%% file: paper.tex
\documentclass{lmcs}
\theoremstyle{definition}\newtheorem{definition}[thm]{Definition}
\theoremstyle{definition}\newtheorem{example}[thm]{Example}
\theoremstyle{definition}\newtheorem{nproposal}[thm]{Proposal}
\theoremstyle{plain}\newtheorem{theorem}[thm]{Theorem}

\usepackage{generic}
\usepackage{macros}

\begin{document} 
\title[On Thin Air Reads]{On Thin Air Reads: Towards an Event Structures Model of Relaxed Memory}
\author{Alan Jeffrey}	
\address{Mozilla Research}	

\author{James Riely}	
\address{DePaul University}	

\titlecomment{An earlier version of this paper appeared in LICS2016.}
\begin{abstract}\input{abstract}\end{abstract}
\maketitle

\input{intro}
\input{es}

\input{memory}

\input{drf}
\input{invariants}

\input{fencing}
\input{outro}

\section*{Acknowledgement}
Riely was supported by the National Science Foundation under Grant
No.~1617175.  Any opinions, findings, and conclusions or recommendations
expressed in this material are those of the author and do not necessarily
reflect the views of the National Science Foundation.

The paper have benefited from discussion with JMM working group and the
anonymous referees.

In an earlier version of this paper, Definition~\ref{def:acyclic} of {acyclic
  justification} quantified over $d \in D$ rather than $d \in D\setminus C$.
The new definition was suggested by Simon Cooksey and Mark Batty by email in
April, 2019.  They noted that Example~\ref{ex:p3} failed under the original
definition, since the player configuration
\begin{math}
  C_3
\end{math}
is not acyclicly justified by 
\begin{math}
  C_2, 
\end{math}
which is a valid opponent move.  

\bibliographystyle{abbrvnat}
\bibliography{bib}

\end{document}

%% file: abstract.tex
To model relaxed memory, we propose \emph{confusion-free event structures} over
an alphabet with a \emph{justification} relation.  
Executions are modeled by \emph{justified} configurations,
where every read event has a justifying write event.
Justification alone is too weak a criterion, since it allows cycles
of the kind that result in so-called thin-air reads. 
\emph{Acyclic justification} forbids such cycles, but also invalidates
event reorderings that result from compiler optimizations and dynamic
instruction scheduling.  
We propose the notion of
\emph{well-justification}, based on a game-like model, which strikes a middle ground.

We show that well-justified configurations satisfy the \emph{DRF theorem}: in
any data-race free program, all well-justified configurations are
sequentially consistent.
We also show that \emph{rely-guarantee reasoning} is sound for well-justified
configurations, but not for justified configurations. For example,
well-justified configurations are \emph{type-safe}. 


Well-justification allows many, but not all reorderings performed by
relaxed memory.  In particular, it fails to validate the commutation of independent
reads.  We discuss variations that may address these
shortcomings.


%% file: intro.tex
\section{Introduction}
\label{sec:intro}


The last few decades have seen several attempts to define a suitable
semantics for shared-memory concurrency under relaxed assumptions;
see \citet{DBLP:conf/esop/BattyMNPS15} for a recent summary.
\emph{Event structures} \citep{DBLP:conf/ac/Winskel86} provide a way
to visualize all of the conflicting executions of a program as a
single semantic object.  In this paper, we exploit the visual nature
of event structures to provide a fresh approach to relaxed memory
models.

Consider a simple programming language where all values are booleans,
registers (ranged over by \texttt{r}) are thread-local and variables
(ranged over by \texttt{x} and \texttt{y}) are global.  In order to define the semantics
compositionally, variable read is defined as a choice among the possible
values that might be read.
For example,
the event structure for \texttt{(r=x; y=r;)} is as follows.
\begin{gather*}
  \begin{tikzpicture}[node distance=1em]
    \zevent{init}{\init}{}
    \zevent{x0}{\DR{x}{0}}{below left=of init}
    \zevent{x1}{\DR{x}{1}}{below right=of init}
    \zevent{y0}{\DW{y}{0}}{below=of x0}
    \zevent{y1}{\DW{y}{1}}{below=of x1}
    \zpo{init}{x0}
    \zpo{init}{x1}
    \zpo{x0}{y0}
    \zpo{x1}{y1}
    \zconfl{x0}{x1}
  \end{tikzpicture}
\end{gather*}
Register values are resolved via substitution and therefore do not appear in
the event structure.  The arrows represent program order, and the zigzag
represents a primitive conflict.  If two events are in conflict, then all following
events are also in conflict.

This structure has two maximal conflict-free \emph{configurations}, which represent
a possible execution of the program:
\begin{align*}
  &\begin{tikzpicture}[node distance=1em]
    \zevent{init}{\init}{}
    \zevent{x0}{\DR{x}{0}}{right=of init}
    \zevent{y0}{\DW{y}{0}}{right=of x0}
    \zrwj[out=40,in=140]{init}{x0};
    \zpo{init}{x0}
    \zpo{x0}{y0}
  \end{tikzpicture}
  \text{~~and}
  &\begin{tikzpicture}[node distance=1em]
    \zevent{init}{\init}{}
    \zevent{x1}{\DR{x}{1}}{right=of init}
    \zevent{y1}{\DW{y}{1}}{right=of x1}
    \zpo{init}{x1}
    \zpo{x1}{y1}
  \end{tikzpicture}.
\end{align*}
If we suppose that this code fragment is embedded in a larger program, the
two configurations are equally sensible: $x$ could be anything.  However, if
we take $\init$ to be the top-level initialization of the program and suppose
that variables are initialized to $0$, then the first configuration above
seems sensible, whereas the second does not: $x$ must be $0$.

A read event is \emph{justified} by a matching \emph{visible} write, drawn with a dashed arrow in the above configurations.
Writes are \emph{hidden} if they occur later or are blocked by an
intervening write.  When modeling
executions of whole programs, one expects that all reads in a
configuration must be justified.

In a happens-before model \citep{DBLP:conf/popl/MansonPA05}, all concurrent writes are visible, 
making this notion of justification quite permissive.  Consider a
program with two threads
and the corresponding event structure, with events numbered for reference.
\begin{gather*}
  \tag{\ensuremath{P_1}} \label{eq:p1}
  \texttt{(\zr1=x; y=\zr1;) || (\zr2=y; x=\zr2;)}
  \\
  \begin{tikzpicture}[node distance=1em,baselinecenter]
    \zeventr{10}{init}{\init}{}
    \zeventr{12}{Rx1}{\DR{x}{1}}{below left=of init}
    \zeventl{11}{Rx0}{\DR{x}{0}}{left of=Rx1,node distance=5em}
    \zeventl{15}{Wy0}{\DW{y}{0}}{below=of Rx0}
    \zeventr{16}{Wy1}{\DW{y}{1}}{below=of Rx1}
    \zeventl{13}{Ry0}{\DR{y}{0}}{below right=of init}
    \zeventr{14}{Ry1}{\DR{y}{1}}{right of=Ry0,node distance=5em}
    \zeventl{17}{Wx0}{\DW{x}{0}}{below=of Ry0}
    \zeventr{18}{Wx1}{\DW{x}{1}}{below=of Ry1}
    \zpo{init}{Rx0}
    \zpo{init}{Rx1}
    \zpo{Rx0}{Wy0}
    \zpo{Rx1}{Wy1}
    \zpo{init}{Ry0}
    \zpo{init}{Ry1}
    \zpo{Ry0}{Wx0}
    \zpo{Ry1}{Wx1}
    \zconfl{Rx0}{Rx1}
    \zconfl{Ry0}{Ry1}
  \end{tikzpicture}
\end{gather*}
Here, the events that are neither ordered nor in conflict are {concurrent}.
The event structure for \ref{eq:p1} has the following configuration,
in which every read event is justified by a matching write that is either
before it, or concurrent:
\begin{gather*}
  \begin{tikzpicture}[node distance=1em,baselinecenter]
    \tag{\ensuremath{C_0}} \label{eq:c0}
    \zeventr{10}{init}{\init}{}
    \zeventl{11}{Rx0}{\DR{x}{0}}{below left=of init}
    \zeventl{15}{Wy0}{\DW{y}{0}}{below=of Rx0}
    \zeventr{13}{Ry0}{\DR{y}{0}}{below right=of init}
    \zeventr{17}{Wx0}{\DW{x}{0}}{below=of Ry0}
    \zpo{init}{Rx0}
    \zpo{Rx0}{Wy0}
    \zpo{init}{Ry0}
    \zpo{Ry0}{Wx0}
    \zrwj{Wy0}{Ry0}
    \zrwj[out=180,in=90]{init}{Rx0}
  \end{tikzpicture}
\end{gather*}
Unfortunately, the event structure also has a configuration
in which there is a cycle in justification-and-program-order:
\begin{gather*}
  \begin{tikzpicture}[node distance=1em,baselinecenter]
    \tag{\ensuremath{C_1}} \label{eq:c1}
    \zeventr{10}{init}{\init}{}
    \zeventl{12}{Rx1}{\DR{x}{1}}{below left=of init}
    \zeventl{16}{Wy1}{\DW{y}{1}}{below=of Rx1}
    \zeventr{14}{Ry1}{\DR{y}{1}}{below right=of init}
    \zeventr{18}{Wx1}{\DW{x}{1}}{below=of Ry1}
    \zpo{init}{Rx1}
    \zpo{Rx1}{Wy1}
    \zpo{init}{Ry1}
    \zpo{Ry1}{Wx1}
    \zrwj{Wy1}{Ry1}
    \zrwj{Wx1}{Rx1}
  \end{tikzpicture}
\end{gather*}
Due to the cycle, any available value can be so justified, thus arising ``out
of thin air''.  Some memory models have undefined semantics in the presence
of such data races \citep{DBLP:conf/popl/BattyOSSW11}.
In the absence of such undefined behaviors, however,
languages that claim
memory safety must disallow thin-air values in order to preserve
type safety.

Unfortunately, cycles such as those in configuration \ref{eq:c1} cannot be
banned outright without also banning useful program transformations, such as
instruction reordering.  For example, consider the following program.
\begin{gather}
  \tag{\ensuremath{P_2}} \label{eq:p2}
  \texttt{(\zr1=x; y=1;) || (\zr2=y; x=1;)}
\end{gather}
The event structure for \ref{eq:p2} is the same as that for \ref{eq:p1}
except that all writes have value $1$.  Thus, \ref{eq:p2} also allows
configuration \ref{eq:c1}.  Clearly, if the order of the two instructions is
swapped in either thread of \ref{eq:p2}, then it is possible for both threads
to read $1$.  Since program transformations may not introduce new behaviors,
\ref{eq:c1} must also be considered a valid configuration of the original
program.

There are several models in the literature designed to allow configuration
\ref{eq:c1} for \ref{eq:p2}, yet deny it for \ref{eq:p1}.  Roughly these can
be divided into two approaches: working with multiple executions
\citep{DBLP:conf/popl/MansonPA05,2010-gosrmm} or working with axioms and
rewrite rules
\citep{DBLP:conf/esop/CenciarelliKS07,Saraswat:2007:TMM:1229428.1229469,Sewell16}.

We propose a new approach, based on two-player games.  The game is as
follows: we start in configuration $C$, and the player's goal is to
extend it to configuration $D$.  The opponent picks a configuration $C'$
which includes $C$, and whose new events are acyclically justified.
The player then picks a configuration $C''$ which includes $C'$, and
whose new events are also acyclically justified. If $C''$ justifies $D$
then the player has won, otherwise the opponent has won. If the player
has a winning strategy for this game, we say that $C$ \emph{Always Eventually
  (AE) justifies} $D$.

From this game, we can define the \emph{well-justified} configurations
inductively: $\emptyset$ is well-justified; if $C$ is well-justified
and $C$ AE-justifies $D$ then $D$ is well-justified.

Consider the following program, \ref{eq:p3}.
\begin{gather*}
  \tag{\ensuremath{P_3}} \label{eq:p3}
  \texttt{\zr1=x; y=1; || \zr2=y; x=\zr2;}
  \\[1ex]
  \begin{tikzpicture}[node distance=1em,baselinecenter]
    \zeventr{30}{init}{\init}{}
    \zeventr{32}{Rx1}{\DR{x}{1}}{below left=of init}
    \zeventl{31}{Rx0}{\DR{x}{0}}{left of=Rx1,node distance=5em}
    \zeventl{35}{Wy0}{\DW{y}{1}}{below=of Rx0}
    \zeventr{36}{Wy1}{\DW{y}{1}}{below=of Rx1}
    \zeventl{33}{Ry0}{\DR{y}{0}}{below right=of init}
    \zeventr{34}{Ry1}{\DR{y}{1}}{right of=Ry0,node distance=5em}
    \zeventl{37}{Wx0}{\DW{x}{0}}{below=of Ry0}
    \zeventr{38}{Wx1}{\DW{x}{1}}{below=of Ry1}
    \zpo{init}{Rx0}
    \zpo{init}{Rx1}
    \zpo{Rx0}{Wy0}
    \zpo{Rx1}{Wy1}
    \zpo{init}{Ry0}
    \zpo{init}{Ry1}
    \zpo{Ry0}{Wx0}
    \zpo{Ry1}{Wx1}
    \zconfl{Rx0}{Rx1}
    \zconfl{Ry0}{Ry1}
  \end{tikzpicture}
\end{gather*}
We show that both reads may be resolved to $1$ in the well-justified
configuration $\set{\zz{30}\C\zz{32}\C\zz{34}\C\zz{36}\C\zz{38}}$.
In this case the cyclic justifier models a valid execution,
caused by a compiler or hardware optimization reordering
\texttt{(\zr1=x; y=1;)} as \texttt{(y=1; \zr1=x;)}.

We first show that $\emptyset$ AE-justifies $\set{\zz{30}\C\zz{34}\C\zz{38}}$.
The opponent may choose any configuration acyclically
justified from $\emptyset$; the interesting choices are the maximal configurations
\begin{math}
  \set{\zz{30}\C\zz{31}\C\zz{33}\C\zz{35}\C\zz{37}} \text{ and }
  \set{\zz{30}\C\zz{31}\C\zz{34}\C\zz{35}\C\zz{38}}.
\end{math}
Since both of these include $\zz{35}$, which justifies $\zz{34}$, the player
does not have to add any events to justify $\set{\zz{30}\C\zz{34}\C\zz{38}}$.
Note that $\emptyset$ does \emph{not} AE-justify $\set{\zz{30}\C\zz{32}}$,
since the opponent can choose the configuration
\begin{math}
  \set{\zz{30}\C\zz{31}\C\zz{35}\C\zz{33}\C\zz{37}}.
\end{math}

We now show that the configuration $\set{\zz{30}\C\zz{34}\C\zz{38}}$ AE-justifies
$\set{\zz{30}\C\zz{32}\C\zz{34}\C\zz{36}\C\zz{38}}$.  The opponent may
choose any configuration acyclically justified from
$\set{\zz{30}\C\zz{34}\C\zz{38}}$; since any choice includes
$\zz{38}$, which justifies $\zz{32}$, the player does not have to add
any events to justify $\set{\zz{30}\C\zz{32}\C\zz{34}\C\zz{36}\C\zz{38}}$.
We have thus shown a cyclic configuration similar to \ref{eq:c1} is 
well-justified for \ref{eq:p3}.
\begin{gather*}
  \begin{tikzpicture}[node distance=1em,baselinecenter]
    \zeventr{30}{init}{\init}{}
    \zeventl{32}{Rx1}{\DR{x}{1}}{below left=of init}
    \zeventl{36}{Wy1}{\DW{y}{1}}{below=of Rx1}
    \zeventr{34}{Ry1}{\DR{y}{1}}{below right=of init}
    \zeventr{38}{Wx1}{\DW{x}{1}}{below=of Ry1}
    \zpo{init}{Rx1}
    \zpo{Rx1}{Wy1}
    \zpo{init}{Ry1}
    \zpo{Ry1}{Wx1}
    \zrwj{Wy1}{Ry1}
    \zrwj{Wx1}{Rx1}
  \end{tikzpicture}
\end{gather*}

This reasoning fails for configuration \ref{eq:c1} of \ref{eq:p1}.  In this
case, the player is unable to establish that $\emptyset$ \AEjustifies{}
$\set{\zz{10}\C\zz{14}\C\zz{18}}$.  We provide a proof in
\textsection\ref{sec:invariants}.  Intuitively, the only maximal
configuration available to the opponent is
$\set{\zz{10}\C\zz{11}\C\zz{13}\C\zz{15}\C\zz{17}}$, and this fails to
justify $\zz{14}$ since there is no write of $1$ to $y$.

\medskip

We review the literature on \emph{confusion-free event structures} in
\textsection\ref{sec:es}.  In \textsection\ref{sec:memory} we define well-justification
and provide further examples.  We give the definition for a
Java-like happens-before model \citep{DBLP:conf/popl/MansonPA05}.
We discuss synchronization actions, such as locks, in \textsection\ref{sec:fencing}.

Perhaps the most important property of a relaxed memory model is \emph{DRF}: that
programs without data races behave as they would with strong memory---that
is, as they would with sequentially consistent memory \citep{Lamport}.  In
\textsection\ref{sec:drf}, we describe our proof of the DRF theorem, which we
have verified in Agda.

In \textsection\ref{sec:invariants}, we show that invariant reasoning is
possible using our definition.  We state a general theorem---also verified in
Agda---which is sufficient to establish type safety for static allocation.

We
describe some of the limitations of our definition in
\textsection\ref{sec:outro}.  While the definition presented here is a step
in the right direction, it fails to validate common reorderings, such as the
reordering of reads on different variables.  We give an alternative
definition that is better behaved on the Java Memory Model causality test
cases \citep{PughWebsite}.  The induction principle used in our proof of DRF
fails for this alternative definition.  

The paper ends with a discussion of related work and open problems.





The Agda development for this paper is available 
\url{https://perma.cc/WAC3-JLXV}.


%% file: es.tex
\section{Event Structures}
\label{sec:es}

Event structures were introduced by
\citet{Winskel:1988:IES:648140.749805} as a non-interleaving model of
concurrency. They are notable for providing a compact model of concurrent
systems, for example an event structure model for $n$ concurrent processes
will often have only $O(n)$ events, compared to the $O(2^n)$ states in a
labeled transition system.

In this section, we review the definitions
associated with conflict-free labeled event structures,
and their visualization as graphs. Readers familiar with
event structures can skip to \textsection\ref{sec:memory}, where
the new material begins.

A \emph{partial order} $(E,{\le})$ is a set $E$
(the \emph{event set}) equipped with a reflexive,
transitive, antisymmetric relation ${\le}$ (the
\emph{causal order}).
A \emph{well order} is a partial order that has no
\begin{wrapfigure}{r}{0.18\columnwidth}
  \centering
  \begin{tikzpicture}[node distance=.6em]
    \seventr0{n0}{}{}
    \seventl1{n1}{}{below left=of n0}
    \seventl2{n2}{}{below=of n1}
    \seventr3{n3}{}{below right=of n0}
    \zpo{n0}{n1}
    \zpo{n1}{n2}
    \zpo{n0}{n3}
  \end{tikzpicture}
\end{wrapfigure}
infinite decreasing sequence.

We visualize partial orders as directed acyclic graphs where edges denote order.
For example the order on $\{0,1,2,3\}$ where $0\le 1\le 2$ and $0\le 3$
is visualized on the right.

A \emph{prime event structure} $(E,{\le},{\confl})$ is a well order
together with a symmetric relation ${\confl}$ on $E$
(the \emph{conflict} relation),
such that if $c \confl d \le e$ then $c \confl e$.

For any prime event structure, define the \emph{primitive conflict} relation
${\muconfl}$ on $E$ as $d \muconfl e$ whenever $d \confl e$ and for any
$d \ge b \confl c \le e$ we have $d=b$ and $c=e$. Primitive conflict is also
known as \emph{minimal} conflict.  A prime event structure is
\emph{confusion-free}~\citep{Nielsen:1979:PNE:647172.716120} whenever
$\muconfl$ is transitive, and if $c \le d \muconfl e$ then $c \le e$.

For any confusion-free event structure, define the
\emph{primitive conflict equivalence} $d \muconfleq e$
whenever $d=e$ or $d \muconfl e$. It is routine to show
that primitive conflict equivalence is symmetric and transitive,
and hence forms an equivalence 
on $E$.

\begin{wrapfigure}{r}{0.18\columnwidth}
  \centering
  \begin{tikzpicture}[node distance=.6em]
    \seventr0{n0}{}{}
    \seventl1{n1}{}{below left=of n0}
    \seventl2{n2}{}{below=of n1}
    \seventr3{n3}{}{below right=of n0}
    \zpo{n0}{n1}
    \zpo{n1}{n2}
    \zpo{n0}{n3}
    \zconfl{n1}{n3}
  \end{tikzpicture}
\end{wrapfigure}
We visualize confusion-free event structures by including the
primitive conflict equivalence in the visualization. For example
the event structure which extends the previous partial order with
$1 \confl 3$ and $2 \confl 3$ has $1 \muconfleq 3$, so is visualized as on
the right.

A \emph{labeled event structure} $(E,{\le},{\confl},\lambda)$
over a \emph{label set} $\Sigma$
is a prime event structure together with a function $\lambda:E\fun\Sigma$.

We visualize labeled event structures as node-labeled graphs.
For example the labeled event structure which extends the previous
event structure with labeling $\lambda(0)=\init$, $\lambda(1)=\DR{x}{0}$,
$\lambda(2)=\DW{y}{1}$ and $\lambda(3)=\DR{x}{1}$ is visualized as follows.
\begin{displaymath}
  \begin{tikzpicture}[node distance=1em]
    \zevent{n0}{\init}{}
    \zevent{n1}{\DR{x}{0}}{below left=of n0}
    \zevent{n2}{\DW{y}{1}}{below=of n1}
    \zevent{n3}{\DR{x}{1}}{below right=of n0}
    \zpo{n0}{n1}
    \zpo{n1}{n2}
    \zpo{n0}{n3}
    \zconfl{n1}{n3}
  \end{tikzpicture}
\end{displaymath}

For any prime event structure, a set $C \subseteq E$
is \emph{conflict-free} whenever there is no $d,e \in C$
such that $d \confl e$. $C$ is \emph{down-closed}
whenever $d \le e \in C$ implies $d \in C$.
A \emph{configuration} is a set which is
conflict-free and down-closed.

Since configurations are conflict-free, they can be visualized
as node-labeled directed acyclic graphs, for example the
two largest configurations for the previous labeled event structure
are
\begin{displaymath}
  \begin{tikzpicture}[node distance=1em]
    \zevent{n0}{\init}{}
    \zevent{n1}{\DR{x}{0}}{right=of n0}
    \zevent{n2}{\DW{y}{1}}{right=of n1}
    \zpo{n0}{n1}
    \zpo{n1}{n2}
  \end{tikzpicture}
  \quad\mbox{and}\quad
  \begin{tikzpicture}[node distance=1em]
    \zevent{n0}{\init}{}
    \zevent{n3}{\DR{x}{1}}{right=of n0}
    \zpo{n0}{n3}
  \end{tikzpicture}.
\end{displaymath}

Given labeled event structures $\ES_1= (E_1,{\le}_1,{\confl}_1,\lambda_1)$ and $\ES_2= (E_2,{\le}_2,{\confl}_2,\lambda_2)$
(without loss of generality, we assume event sets $\ES_1$ and $\ES_2$ are disjoint)
define the \emph{parallel composite} event structure $\ES_1\epar\ES_2$ as having:
\begin{itemize}

\item event set $E$ is $E_1 \cup E_2$,
\item causal order $\le$ is ${\le_1} \cup {\le_2}$,
\item conflict $\confl$ is ${\confl_1} \cup {\confl_2}$, and
\item labeling $\lambda$ is $\lambda_1 \cup \lambda_2$.
  
\end{itemize}
The \emph{sum} event structure $\ES_1+\ES_2$ is the same except:
\begin{itemize}

\item conflict $\confl$ is ${\confl_1} \cup {\confl_2}
  \cup (E_1\times E_2) \cup (E_2 \times E_1)$.
  
\end{itemize}
We write $0$ for the empty event structure with event set $\emptyset$.

For a label $\sigma\in\Sigma$, the prefix $\esprefix{\sigma}{\ES_0}$ introduces a new $\sigma$-labeled
event ordered before all the events of $\ES_0$. It is defined as having:
\begin{itemize}

\item event set $E$ is $E_0 \cup \{\bot\}$,

\item causal order $\le$ is ${\le_0} \cup (\{\bot\}\times E)$,
\item conflict $\confl$ is $\confl_0$, and
\item labeling $\lambda$ is $\lambda_0 \cup \{(\bot,\sigma)\}$.

\end{itemize}


Using an appropriate alphabet (discussed in more detail in
\textsection\ref{sec:memory}), we can give the semantics of a simple shared-memory
concurrent language.  The construction uses sum, parallel composition, prefix and the
empty event structure.



Let $r$ range over \emph{registers}.  A \emph{store} maps registers to
values.  Let $\rho$ range over stores and $\rho_0$ be the initial store,
which maps all registers to $0$.  We write $\update\rho{r}{v}$ for store
update:
\begin{displaymath}
  \update\rho{r}{v}(r') =
  \begin{cases}
    v &\textif r=r'\\
    \rho(r') &\textotherwise
  \end{cases}
\end{displaymath}
Let $M$ range over \emph{expressions}, which may include registers, but
not variables.  Let $V$ be a set of \emph{values}.  Let $\M{\cdot}$ be an
interpretation that maps expressions and stores to values.

We give the semantics of a single-threaded program,
featuring reads, writes and conditionals as an event
structure:
\begin{alignat*}{2}
  \T{ \pget{r}{x} T } ρ &\eqdef \textstyle\sum_{v\in V} \esprefix{\DR{x}{v}}{\T{T}{(\update{ρ}{r}{v})}} \\
  \T{ \pset{x}{M} T } ρ &\eqdef \esprefix{\DW{x}{\M{M}{ρ}}}{\T{T}{ρ}} \\
  \T{ \pdone } ρ &\eqdef 0 \\
  \T{ \pif{M}{T_1}{T_0} } ρ &\eqdef 
  \begin{cases}
    \T{T_0}{ρ} & \textif \M{M}{ρ} = 0 \\
    \T{T_1}{ρ} & \textotherwise
  \end{cases}
\end{alignat*}
A \emph{program} is an collection of threads $T_1 \ppar\cdots\ppar T_n$,
interpreted as parallel composition of event structures with an initial event:
\begin{displaymath}
  \P{T_1 \ppar\cdots\ppar T_n} \eqdef \esprefix{\init}{(\T{T_1}{ρ₀} \epar\cdots\epar \T{T_n}{ρ₀})}
\end{displaymath}
We use standard abbreviations.
For example, the program
\begin{displaymath}
  \texttt{if (x==0) \{y=1;\}}
\end{displaymath}
desugars to the following.
\begin{displaymath}
  \texttt{r=x; if(r==0)\{y=1; done\}\,else\,\{done\}}
\end{displaymath}
If we take $V=\set{0\C 1}$, then this has semantics
\begin{displaymath}
  \esprefix{\init}{(%
    (\esprefix{\DR{x}{0}}{\esprefix{\DW{y}{1}}{0}})
    +
    (\esprefix{\DR{x}{1}}{0})
    )}
\end{displaymath}
visualized as follows.
\begin{displaymath}
  \begin{tikzpicture}[node distance=1em]
    \zevent{n0}{\init}{}
    \zevent{n1}{\DR{x}{0}}{below left=of n0}
    \zevent{n2}{\DW{y}{1}}{below=of n1}
    \zevent{n3}{\DR{x}{1}}{below right=of n0}
    \zpo{n0}{n1}
    \zpo{n1}{n2}
    \zpo{n0}{n3}
    \zconfl{n1}{n3}
  \end{tikzpicture}
\end{displaymath}

Note that in this semantics, conflict is only introduced by reads.  Each
conflicting event represents the read of a distinct value; since only one
value can be read, the events are in primitive conflict.


%% file: memory.tex
\section{Memory Event Structures}
\label{sec:memory}

A \emph{memory alphabet} $(\Sigma,R,W,J,K)$ consists of
\begin{itemize}
\item a set $\Sigma$ (the \emph{actions}),
\item set $R\subseteq\Sigma$ (the \emph{read actions})
\item set $W\subseteq\Sigma$ (the \emph{write actions}),
\item binary relation $J\subseteq(W\times R)$
  (\emph{justification}), and
\item binary relation $K \subseteq J$ (\emph{synchronized justification}).
\end{itemize}
When $(a,b)\in J$, we say that $a$ \emph{justifies} $b$.
Synchronization does not play a role in this section;
we return to it in \textsection\ref{sec:fencing}.

A \emph{memory event structure} over such a memory alphabet is
a confusion-free labeled event structure over $\Sigma$.

The prototypical memory alphabet consists of an initial action, and
read and write actions over some set of \emph{variables} $X$, and some
set of \emph{values} $V$:
\begin{alignat*}{2}
  \Sigma & = R \cup W \\
  R & = \{ \DR{x}{v} \mid x \in X, v \in V \} \\
  W & = \{ \DW{x}{v} \mid x \in X, v \in V \} \cup \{ \init \}
\end{alignat*}
In \textsection\ref{sec:es} we saw that such an alphabet can be used to give the
semantics for a simple shared-memory concurrent language. The justification
relation for this alphabet is that $\init$ justifies a read of $0$,
and that a write of $v$ justifies a read of $v$ to the same variable:
\begin{alignat*}{2}
  J & = \{ (\init,\DR{x}{0}) \mid x \in X \} \\
    &\cup \{ (\DW{x}{v},\DR{x}{v}) \mid x \in X, v \in V \}
\end{alignat*}

In a memory event structure, an event $e$
is a \emph{read event} whenever $\lambda(e)\in R$, 
and a \emph{write event} whenever $\lambda(e)\in W$.
The sets $R$ and $W$ need not be disjoint; thus, a memory alphabet may include
read-modify-write actions such as exchange, compare-and-set or
increment.  The semantics of these operators as event structures is straightforward, following
the style given in \textsection\ref{sec:es}.

In a memory event structure, we can lift justification from labels
to events, but this is not just a matter of looking at the labeling,
since events should not be justified by later events, by events in conflict,
or by events with an intervening event in read-write conflict. For example, 
in the program
\begin{displaymath}
  \texttt{if(y)\{x=0;\}\,else\,\{x=1;x=x;\}}
\end{displaymath}
has the following event structure semantics, where we visualize justification
as a dashed edge.
\begin{displaymath}
  \begin{tikzpicture}[node distance=1em]
    \zevent{init}{\init}{}
    \zevent{Ry0}{\DR{y}{0}}{below left=of init}
    \zevent{Ry1}{\DR{y}{1}}{below right=of init}
    \zevent{Wx0b}{\DW{x}{0}}{below=of Ry1}
    \zevent{Wx1}{\DW{x}{1}}{below=of Ry0}
    \zevent{Rx0}{\DR{x}{0}}{below left=of Wx1}
    \zevent{Rx1}{\DR{x}{1}}{below right=of Wx1}
    \zevent{Wx0c}{\DW{x}{0}}{below=of Rx0}
    \zevent{Wx1c}{\DW{x}{1}}{below=of Rx1}
    \zpo{init}{Ry0}
    \zpo{init}{Ry1}
    \zpo{Ry1}{Wx0b}
    \zpo{Ry0}{Wx1}
    \zpo{Wx1}{Rx0}
    \zpo{Wx1}{Rx1}
    \zpo{Rx0}{Wx0c}
    \zpo{Rx1}{Wx1c}
    \zconfl{Rx0}{Rx1}
    \zconfl{Ry0}{Ry1}
    \zrwj[out=180,in=90]{init}{Ry0}
    \zrwj[out=0,in=90]{Wx1}{Rx1}
  \end{tikzpicture}
\end{displaymath}
There is no event labeled $\DW{y}{1}$; therefore, the only justified read of
$y$ is 0, not 1.  Neither event labeled $\DW{x}{0}$ justifies $\DR{x}{0}$:
one is in conflict with it and the other is later.  Finally, $\init$ does not justify $\DR{x}{0}$
because $\DW{x}{1}$ is an intervening event in read-write conflict.  Thus,
the only justified read of $x$ is 1, not 0.

In a memory event structure, we say write event $d$ \emph{justifies} read
event $e$ whenever:
\begin{enumerate}
\item $(\lambda(d),\lambda(e)) \in J$,
\item we do not have $e < d$,
\item we do not have $d \confl e$, and
\item there is no $d < b < c \muconfleq e$
  such that $(\lambda(b),\lambda(c)) \in J$.
  \label{item:just-noblock}
\end{enumerate}
Visually these conditions are that $d$ cannot justify $e$
when:
\begin{displaymath}
  \begin{tikzpicture}[node distance=1em,baselinecenter]
    \zevent{d}{d}{}
    \zevent{e}{e}{above=of d}
    \zpo{e}{d}
  \end{tikzpicture}
  \quad\textor\quad
  \begin{tikzpicture}[node distance=1em,baselinecenter]
    \zevent{b}{}{}
    \zevent{c}{}{right=of b}
    \zevent{d}{d}{below=of b}
    \zevent{e}{e}{below=of c}
    \zpo{b}{d}
    \zpo{c}{e}
    \zconfl{b}{c}
  \end{tikzpicture}
  \quad\textor
  \begin{tikzpicture}[node distance=1em,baselinecenter]
    \zevent{d}{d}{}
    \zevent{b}{b}{below=of d}
    \zevent{c}{c}{below left=of b}
    \zevent{e}{e}{below right=of b}
    \zpo{d}{b}
    \zpo{b}{c}
    \zconfl{c}{e}
    \zrwj[out=180,in=90]{b}{c}
  \end{tikzpicture}
\end{displaymath}
Item~(\ref{item:just-noblock}) provides a formal definition of what it means
for $b$ to be in \emph{read-write conflict} with $e$.
The following statement is equivalent:
``there is no $d < b < e \muconfleq c$
such that $b$ justifies $c$.''
However, we cannot use this as the definition without worrying about
circularity.

\begin{definition}[Justified]
  A configuration $C$ is \emph{justified} whenever every read event in $C$ is
  justified by at least one write event in $C$. \qed
\end{definition}

For example, the program \texttt{y=x;} has two maximal configurations, but
only one of them is justified:
\begin{align*}
  \begin{tikzpicture}[node distance=1em,baselinecenter]
    \zevent{init}{\init}{}
    \zevent{x0}{\DR{x}{0}}{below=of init}
    \zevent{y0}{\DW{y}{0}}{below=of x0}
    \zrwj[out=180,in=180]{init.west}{x0.west};
    \zpo{init}{x0}
    \zpo{x0}{y0}
  \end{tikzpicture}
  \quad\textand\quad
  \begin{tikzpicture}[node distance=1em,baselinecenter]
    \zevent{init}{\init}{}
    \zevent{x1}{\DR{x}{1}}{below=of init}
    \zevent{y1}{\DW{y}{1}}{below=of x1}
    \zpo{init}{x1}
    \zpo{x1}{y1}
  \end{tikzpicture}
\end{align*}

An event may have multiple justifiers in a single configuration.  In
drawings, we choose one.  For example, the program
$\texttt{(x=0; || r=x;)}$ has only one maximal configuration.  The
read event in this configuration has two justifiers: 
\begin{displaymath}
  \begin{tikzpicture}[node distance=1em,baselinecenter]
    \zeventr{}{init}{\init}{}
    \zeventl{}{Wx0}{\DW{x}{0}}{below left=of init}
    \zeventr{}{Rx0}{\DR{x}{0}}{below right=of init}
    \zpo{init}{Wx0}
    \zpo{init}{Rx0}
    \zrwj[out=0,in=90]{init}{Rx0}
  \end{tikzpicture}
  \qquad
  \begin{tikzpicture}[node distance=1em,baselinecenter]
    \zeventr{}{init}{\init}{}
    \zeventl{}{Wx0}{\DW{x}{0}}{below left=of init}
    \zeventr{}{Rx0}{\DR{x}{0}}{below right=of init}
    \zpo{init}{Wx0}
    \zpo{init}{Rx0}
    \zrwj{Wx0}{Rx0}
  \end{tikzpicture}  
\end{displaymath}

Unfortunately, justified configurations, although necessary, are not a
sufficient condition for modeling valid executions, as they allow cycles in
the union of causal order and justification, which cause \emph{thin air
  reads}.  For example, the program \ref{eq:p1} from the introduction 
includes the justified configurations
\begin{math}
  \set{\zz{10}\C\zz{11}\C\zz{13}\C\zz{15}\C\zz{17}}
\end{math}
and
\begin{math}
  \set{\zz{10}\C\zz{12}\C\zz{14}\C\zz{16}\C\zz{18}}:
\end{math}
\begin{gather*}
  \texttt{(r1=x; y=r1;) || (r2=y; x=r2;)}
  \\
  \begin{tikzpicture}[node distance=1em,baselinecenter]
    \zeventr{10}{init}{\init}{}
    \zeventl{11}{Rx0}{\DR{x}{0}}{below left=of init}
    \zeventl{15}{Wy0}{\DW{y}{0}}{below=of Rx0}
    \zeventr{13}{Ry0}{\DR{y}{0}}{below right=of init}
    \zeventr{17}{Wx0}{\DW{x}{0}}{below=of Ry0}
    \zpo{init}{Rx0}
    \zpo{Rx0}{Wy0}
    \zpo{init}{Ry0}
    \zpo{Ry0}{Wx0}
    \zrwj[out=180,in=90]{init}{Rx0}
    \zrwj[out=0,in=90]{init}{Ry0}
  \end{tikzpicture}
  \quad
  \begin{tikzpicture}[node distance=1em,baselinecenter]
    \zeventr{10}{init}{\init}{}
    \zeventl{12}{Rx1}{\DR{x}{1}}{below left=of init}
    \zeventl{16}{Wy1}{\DW{y}{1}}{below=of Rx1}
    \zeventr{14}{Ry1}{\DR{y}{1}}{below right=of init}
    \zeventr{18}{Wx1}{\DW{x}{1}}{below=of Ry1}
    \zpo{init}{Rx1}
    \zpo{Rx1}{Wy1}
    \zpo{init}{Ry1}
    \zpo{Ry1}{Wx1}
    \zrwj{Wy1}{Ry1}
    \zrwj{Wx1}{Rx1}
  \end{tikzpicture}
\end{gather*}
In the latter, there is a cycle in causal\BreakablePlus{}justification order.
It is straightforward to ban such cycles.  


Recall that a configuration is a subset of the event set $E$.

\begin{definition}[Acyclically justified]
  \label{def:acyclic}
  On configurations, define $C$ \emph{justifies} $D$ whenever for any read
  event $d \in D\setminus C$ there exists a write event $c \in C$ such that $c$
  justifies $d$.

  Write $C \acextone D$ whenever $C \subseteq D$ and $C$ justifies $D$.

  Write $\acext$ for the reflexive, transitive closure of $\acextone$.

  Define $C$ is \emph{acyclically justified} whenever $\emptyset \acext C$. \qed
\end{definition}

A justified configuration is acyclically justified whenever the order induced
by causal order and justification is acyclic.  Any acyclically justified
configuration is also justified.

For example, for \ref{eq:p1}, we have that
\begin{math}
  \emptyset \acextone \set{\zz{10}},
\end{math}
since $\zz{10}$ is not a read.
In addition, we have
\begin{math}
  \set{\zz{10}} \acextone \set{\zz{10}\C\zz{11}}
\end{math}
and
\begin{math}
  \set{\zz{10}} \acextone \set{\zz{10}\C\zz{13}}
\end{math}
since $\zz{10}$ justifies both $\zz{11}$ and $\zz{13}$.  Taking a maximal
configuration at each step, we have:
\begin{displaymath}
  \emptyset\acextone \set{\zz{10}} \acextone 
  \set{\zz{10}\C\zz{11}\C\zz{13}\C\zz{15}\C\zz{17}}
\end{displaymath}
However, there is no such chain leading from $\emptyset$ to
\begin{math}
  \set{\zz{10}\C\zz{12}\C\zz{14}\C\zz{16}\C\zz{18}}.
\end{math}

Consider the following program.
\begin{gather*}
  \tag{\ensuremath{P_4}} \label{eq:p4}
  \texttt{y=x; || x=1 || r=y;}
  \\
        \begin{tikzpicture}[node distance=1em,baselinecenter]
          \zeventr{40}{init}{\init}{}
          \zeventr{42}{Rx1}{\DR{x}{1}}{below left=of init}
          \zeventl{41}{Rx0}{\DR{x}{0}}{left of=Rx1,node distance=5em}
          \zeventl{45}{Wy0}{\DW{y}{0}}{below=of Rx0}
          \zeventr{46}{Wy1}{\DW{y}{1}}{below=of Rx1}
          \zeventl{43}{Ry0}{\DR{y}{0}}{below right=of init}
          \zeventr{44}{Ry1}{\DR{y}{1}}{right of=Ry0,node distance=5em}
          \zeventr{47}{Wx1}{\DW{x}{1}}{below=of Ry0,xshift=-1em}
          \zpo{init}{Rx0}
          \zpo{init}{Rx1}
          \zpo[out=270,in=135]{init}{Wx1}
          \zpo{Rx0}{Wy0}
          \zpo{Rx1}{Wy1}
          \zpo{init}{Ry0}
          \zpo{init}{Ry1}
          \zconfl{Rx0}{Rx1}
          \zconfl{Ry0}{Ry1}
        \end{tikzpicture}
\end{gather*}
In this case the write to $x$ is immediately
available, since it is not causally dependent on any read.  Thus:
\begin{gather*}
  \emptyset\acextone \set{\zz{40}\C\zz{47}} \acextone \set{\zz{40}\C\zz{47}\C\zz{41}\C\zz{45}} \acextone \set{\zz{40}\C\zz{47}\C\zz{41}\C\zz{45}\C\zz{43}}
  \\
  \emptyset\acextone \set{\zz{40}\C\zz{47}} \acextone \set{\zz{40}\C\zz{47}\C\zz{42}\C\zz{46}} \acextone \set{\zz{40}\C\zz{47}\C\zz{42}\C\zz{46}\C\zz{44}}
\end{gather*}
Here the read of $x$ is a coin-toss, which determines whether it is possible
to read $\DR{y}{1}$: A configuration that contains $\zz{41}$ cannot also
contain $\zz{44}$.  

The second of these sequences can be seen in Figure~\ref{fig:p4:ae}, read from
top to bottom.  The events included in each successive configuration are
highlighted using a darker, blue background.  Events that are in conflict
with an included event are covered in white.  Thus in a maximal
configuration, such as the last configuration in Figure~\ref{fig:p4:ae}, all
events are either highlighted or covered.  

Acyclic justification rules out cycles, since in any
acyclically justified $C$, there must be configurations
$\emptyset = C_0 \acextone \cdots \acextone C_n = C$,
and for any read event $e\in C$ there must be a $j$
such that $e\in C_{j+1}$ and a $d \in C_j$ which justifies
$e$. Since configurations are $\le$-closed, this means that
there is no infinite sequence $d_1 \le e_1,d_2\le e_2,\dots,$ 
where $d_i$ justifies $e_{i+1}$, and in particular there are no
cycles.

\input{fig-p4-ae}
 Unfortunately, acyclic justification is too strong
a requirement, as it rules out some
valid executions in the presence of optimizations which
reorder memory accesses. For example, the program 
\begin{math}
  \text{\ref{eq:p3}} = \texttt{(r=x; y=1; || x=y;)}
\end{math}
has the event structure given in the introduction.
In this case the cyclic justifier models a valid execution,
caused by a compiler or hardware optimization reordering
\texttt{(r=x; y=1;)} as \texttt{(y=1; r=x;)}. If we are going to
admit such reorderings, we cannot model valid executions by
a property of configurations, and must look at the entire event
structure (this observation was made, in a different model, by
\citet{DBLP:conf/esop/BattyMNPS15}).

\begin{definition}[Well-justified]
  \label{def:well}
  On configurations, define $C$ \emph{always eventually justifies
    (\AEjustifies)} $D$ whenever for any $C \acext C'$ there exists a
  $C' \acext C''$ such that $C''$ justifies $D$.  

  Write $C \aeextone D$ whenever $C \subseteq D$ and $C$ \AEjustifies{} $D$.

  Write $\aeext$ for the reflexive, transitive closure of $\aeextone$. 

  Define $C$ is \emph{well-justified} whenever $C$ is justified and
  $\emptyset \aeext C$. \qed
\end{definition}

Note the player's choice of $D$ is not required to include the opponents
choices in $C'$.  The definition requires only that the player can find an
extension $C''$ of $C'$ that justifies every event in $D$.

A well-justified configuration must be both justified and \AEjustified{}.
The notion of \AEjustification{} describes when a read event is justified by
some write event no matter which execution path is chosen.
\AEjustification{} has the flavor of a two-player game: in a configuration
$C_i$, the opponent chooses a $C_i \acextone C_i'$, after which the player
chooses a $C_i' \acextone C_i''$ which justifies $C_{i+1}$. If the player can
justify $C_{i+1}$ regardless of the opponent move, then the player wins the
round.  The player well-justifies $C$ if
they can repeat this game to move from the initial configuration $\emptyset$ to the final
configuration $C$.

 Any acyclically justified configuration is \AEjustified.  
\begin{example}
Consider the proof of acyclic justification given in
Figure~\ref{fig:p4:ae}.  
Starting from 
\begin{math}
  C_0=\emptyset, 
\end{math} 
the player follows the proof of acyclic justification to select
\begin{math}
  C_1=\set{\zz{40}\C\zz{47}}, 
\end{math} 
then
\begin{math}
  C_2=\set{\zz{40}\C\zz{47}\C\zz{42}\C\zz{46}}
\end{math}
and finally
\begin{math}
  C_3=\set{\zz{40}\C\zz{47}\C\zz{42}\C\zz{46}\C\zz{44}}.
\end{math}
In each case, the events in $C_i$ are justified by an extension $C''_{i-1}$ of $C_{i-1}$,
regardless of the opponent's choice of $C'_{i-1}$.
For example, the opponent may choose 
\begin{math}
  C_1=\set{\zz{40}\C\zz{47}\C\zz{41}}.
\end{math} 
Even though $\zz{41}$ and $\zz{42}$ are in conflict, the player may choose
\begin{math}
  C_2=\set{\zz{40}\C\zz{47}\C\zz{42}\C\zz{46}};
\end{math}
the read $\DR{x}{1}$ of $\zz{42}$ is justified by $\zz{42}$, which the
opponent cannot remove.

For any opponent move $\emptyset\acext C_0'$, the player must choose
$C_0'\acext C_0''$ so that $C_0''$ justifies $C_1$.  In this case, the
player can always choose $C_0''\supseteq\set{\zz{40}\C\zz{47}}$, since
$\zz{40}$ and $\zz{47}$ conflict with no event and do not require justification.  The first two moves of
the player can be collapsed, choosing
$C_1$ to be
\begin{math}
  \set{\zz{40}\C\zz{47}\C\zz{42}\C\zz{46}}, 
\end{math}
since \zz{42} can be justified regardless of the opponent move.  However, the
last two moves cannot be collapsed.  The player cannot initially select
$\zz{44}$; in this case the opponent would win by choosing 
$C_0'=\set{\zz{40}\C\zz{41}\C\zz{43}\C\zz{45}}$. \qed
\end{example}
\begin{example}
\label{ex:p3}%
\input{fig-p3-ae}%
We now consider the proof that \ref{eq:p3} is well-justified to read all
ones, given in Figure~\ref{fig:p3:ae}.  The previous strategy does not work,
since the goal configuration is not acyclically justified.

The player chooses
\begin{math}
  C_1=\set{\zz{30}},
\end{math} 
\begin{math}
  C_2=\set{\zz{30}\C\zz{34}\C\zz{38}}
\end{math}
and finally
\begin{math}
  C_3=\set{\zz{30}\C\zz{34}\C\zz{38}\C\zz{32}\C\zz{36}}.
\end{math}
As in the previous example, the first two player moves can be collapsed, but
not the last two.  We show the first two player moves separately to make the
opponent choices clear.
The opponent can choose $C_1'$ to include any events except $\zz{32}$ (and
therefore $\zz{36}$); there is no acyclically justified configuration that
includes $\zz{32}$.  For this reason events $\zz{32}$ and $\zz{36}$ are gray
in the top configuration of Figure~\ref{fig:p3:ae}.  
The opponent option to include $\zz{33}\in C_1'$ prevents the player from
selecting $\zz{32}\in C_1$. The $\DR{x}{1}$ cannot be justified in
this case.
However, $\DR{y}{1}$ can be justified regardless of the opponent's
choice.  Thus $\zz{34}$ can be included in $C_1$ (or $C_2$, as shown).

Once the player has won the round including $\zz{34}$, the opponent is no
longer at liberty to include $\zz{33}$---the choice has been made.  Thus the
player may include $\zz{32}$ in the next round. 

This reasoning holds even if the write to $y$ is conditional, as in the
following variation:
\begin{gather*}
  \texttt{\zr1=x; if(\zr1<2)\{y=1;\} || \zr2=y; x=\zr2;}
  \\
  \begin{tikzpicture}[node distance=1em,baselinecenter]
    \zevent{init}{\init}{}
    \zevent{Rx2}{\DR{x}{2}}{below left=of init}
    \zevent{Rx1}{\DR{x}{1}}{left of=Rx2,node distance=5em}
    \zevent{Rx0}{\DR{x}{0}}{left of=Rx1,node distance=5em}
    \zevent{Wy0}{\DW{y}{1}}{below=of Rx0}
    \zevent{Wy1}{\DW{y}{1}}{below=of Rx1}
    \zevent{Ry0}{\DR{y}{0}}{below right=of init}
    \zevent{Ry1}{\DR{y}{1}}{right of=Ry0,node distance=5em}
    \zevent{Ry2}{\DR{y}{2}}{right of=Ry1,node distance=5em}
    \zevent{Wx0}{\DW{x}{0}}{below=of Ry0}
    \zevent{Wx1}{\DW{x}{1}}{below=of Ry1}
    \zevent{Wx2}{\DW{x}{2}}{below=of Ry2}
    \zpo[out=180,in=20]{init}{Rx0}
    \zpo[out=185,in=20]{init}{Rx1}
    \zpo{init}{Rx2}
    \zpo{Rx0}{Wy0}
    \zpo{Rx1}{Wy1}
    \zpo{Ry2}{Wx2}
    \zpo{init}{Ry0}
    \zpo[out=-10,in=160]{init}{Ry1}
    \zpo[out=-5,in=160]{init}{Ry2}    
    \zpo{Ry0}{Wx0}
    \zpo{Ry1}{Wx1}
    \zconfl{Rx0}{Rx1}
    \zconfl{Rx1}{Rx2}
    \zconfl{Ry0}{Ry1}
    \zconfl{Ry1}{Ry2}
  \end{tikzpicture}
\end{gather*}
As before, there is no acyclically justified configuration of the event
structure that reads a value other than $0$ for $x$.  Thus the player can
choose an initially choose a configuration containing $\DR{y}{1}$, and play
proceeds as before.
\qed
\end{example}

\begin{example}
  As noted in the introduction, configuration \ref{eq:c1} of \ref{eq:p1}
  fails to be well-justified.  We provide a proof in
  \textsection\ref{sec:invariants}. Intuitively, the player is unable to select
  $\zz{14}\in C_1$, because the opponent can choose $\zz{11}\in C'_0$. \qed
\end{example}

In the process of revising the Java Memory Model, \citet{PughWebsite}
developed a set of twenty \emph{causality test cases}.  Using hand
calculation, we tested our semantics against nineteen of these cases.  (TC9
is based on the idea that an execution should be allowed if there exists an
augmentation, such as thread inlining, that allows it.  This is a non-goal
for our semantics; therefore, we do not consider TC9.)

Our semantics agrees with sixteen of the test cases and disagrees with three:
TC3, TC7 and TC11.  In \textsection\ref{sec:outro}, we discuss TC7, which
best elucidates the issues.

Also by hand calculation, we found that our semantics gives the desired
results for all examples in \citet[\textsection
4]{DBLP:conf/esop/BattyMNPS15} and all but one in \citet[\textsection
5.3]{SevcikThesis}: redundant-write-after-read-elimination---this
counterexample applies to any sensible non-coherent semantics.

\input{fig-sevcik532}

\begin{example}
  \label{ex:sevcik532}
  We consider the redundant-read-elimination counterexample of
  \citep[\textsection{}5.3.2]{SevcikThesis}, given in
  Figure~\ref{fig:sevcik532}.  TC18 of \citep{PughWebsite} is similar.  The
  question is whether there is a well-justified configuration that includes
  event $\zz{e}$.

  Starting from the empty set, the maximal acyclically justified configurations are 
  \begin{math}
    \set{\zz{a}\C \zz{b}\C \zz{f}\C \zz{d}\C \zz{h}}
  \end{math}
  and
  \begin{math}
    \set{\zz{a}\C \zz{c}\C \zz{g}\C \zz{d}\C \zz{h}},
  \end{math}
  both of which include $\zz{h}$ with label $\DW{x}{1}$.
  Therefore the player can choose
  \begin{math}
    \set{\zz{a}\C \zz{c}\C \zz{g}}.
  \end{math}
  
  Having secured $\zz{g}$ with label $\DW{y}{1}$, the player can subsequently
  add $\zz{e}$, with label $\DR{y}{1}$.  At this point, the player has two
  possible extensions, depending on the choice between $\zz{i}$ and $\zz{j}$.
  Both choices are \emph{\AEjustified}, but only the choice of $\zz{j}$ can be
  extended to a \emph{justified} configuration, which is therefore
  \emph{well-justified}: The \AEjustified{} configuration
  \begin{math}
    \set{\zz{a}\C \zz{c}\C \zz{e}\C \zz{g}\C \zz{i}}
  \end{math}
  cannot be extended to include an event labeled $\DW{x}{1}$, which is
  necessary to justify $\zz{e}$ with label $\DR{x}{1}$.

  This example shows the importance of limiting the power of the opponent at
  each step.  If we were to allow the opponent to choose configurations using
  \AEjustification, rather than acyclic justification, then the opponent
  could choose
  \begin{math}
    \set{\zz{a}\C \zz{c}\C \zz{e}\C \zz{g}\C \zz{i}}
  \end{math}
  to extend the empty set, disabling the players initial choice of 
  \begin{math}
    \set{\zz{a}\C \zz{c}\C \zz{g}}.
  \end{math}
  In this case, one can prevent the attack by limiting the opponent to
  well-justification rather than \AEjustification.  By adding another
  variable, however, one can construct an example in which well-justification
  gives the opponent additional power over acyclic justification.
  \qed
\end{example}

%% file: fig-p4-ae.tex
\begin{figure}
  \centering
  \texttt{y=x; || x=1 || r=y;}
  \\[1ex]
    \begin{tikzpicture}[node distance=1em,baselinecenter]
      \zeventr{40}{init}{\init}{}
      \xeventr{42}{Rx1}{\DR{x}{1}}{below left=of init}
      \xeventl{41}{Rx0}{\DR{x}{0}}{left of=Rx1,node distance=5em}
      \xeventl{45}{Wy0}{\DW{y}{0}}{below=of Rx0}
      \xeventr{46}{Wy1}{\DW{y}{1}}{below=of Rx1}
      \xeventl{43}{Ry0}{\DR{y}{0}}{below right=of init}
      \xeventr{44}{Ry1}{\DR{y}{1}}{right of=Ry0,node distance=5em}
      \zeventr{47}{Wx1}{\DW{x}{1}}{below=of Ry0,xshift=-1em}
      \xpo{init}{Rx0}
      \xpo{init}{Rx1}
      \zpo[out=270,in=135]{init}{Wx1}
      \xpo{Rx0}{Wy0}
      \xpo{Rx1}{Wy1}
      \xpo{init}{Ry0}
      \xpo{init}{Ry1}
      \xconfl{Rx0}{Rx1}
      \xconfl{Ry0}{Ry1}
      \zrwj{Wx1}{Rx1}
      \zrwj[out=180,in=35]{init}{Rx0}
      \zrwj[out=0,in=90]{init}{Ry0}
    \end{tikzpicture}    
    \\[2ex]
    \begin{tikzpicture}[node distance=1em,baselinecenter]
      \zeventr{40}{init}{\init}{}
      \zeventr{42}{Rx1}{\DR{x}{1}}{below left=of init}
      \yeventl{41}{Rx0}{\DR{x}{0}}{left of=Rx1,node distance=5em}
      \yeventl{45}{Wy0}{\DW{y}{0}}{below=of Rx0}
      \zeventr{46}{Wy1}{\DW{y}{1}}{below=of Rx1}
      \xeventl{43}{Ry0}{\DR{y}{0}}{below right=of init}
      \xeventr{44}{Ry1}{\DR{y}{1}}{right of=Ry0,node distance=5em}
      \zeventr{47}{Wx1}{\DW{x}{1}}{below=of Ry0,xshift=-1em}
      \ypo{init}{Rx0}
      \zpo{init}{Rx1}
      \zpo[out=270,in=135]{init}{Wx1}
      \ypo{Rx0}{Wy0}
      \zpo{Rx1}{Wy1}
      \xpo{init}{Ry0}
      \xpo{init}{Ry1}
      \yconfl{Rx0}{Rx1}
      \xconfl{Ry0}{Ry1}
      \zrwj[out=0,in=90]{init}{Ry0}
      \xrwj{Wx1}{Rx1}
      \zrwj[out=340,in=270]{Wy1}{Ry1}
    \end{tikzpicture}    
    \\
    \begin{tikzpicture}[node distance=1em,baselinecenter]
      \zeventr{40}{init}{\init}{}
      \zeventr{42}{Rx1}{\DR{x}{1}}{below left=of init}
      \yeventl{41}{Rx0}{\DR{x}{0}}{left of=Rx1,node distance=5em}
      \yeventl{45}{Wy0}{\DW{y}{0}}{below=of Rx0}
      \zeventr{46}{Wy1}{\DW{y}{1}}{below=of Rx1}
      \yeventl{43}{Ry0}{\DR{y}{0}}{below right=of init}
      \zeventr{44}{Ry1}{\DR{y}{1}}{right of=Ry0,node distance=5em}
      \zeventr{47}{Wx1}{\DW{x}{1}}{below=of Ry0,xshift=-1em}
      \ypo{init}{Rx0}
      \zpo{init}{Rx1}
      \zpo[out=270,in=135]{init}{Wx1}
      \ypo{Rx0}{Wy0}
      \zpo{Rx1}{Wy1}
      \ypo{init}{Ry0}
      \zpo{init}{Ry1}
      \yconfl{Rx0}{Rx1}
      \yconfl{Ry0}{Ry1}
      \xrwj{Wx1}{Rx1}
      \xrwj[out=340,in=270]{Wy1}{Ry1}
    \end{tikzpicture}    
  \caption{Acyclic- and \AEjustification{} for \ref{eq:p4}}
  \label{fig:p4:ae}
\end{figure}


%% file: fig-p3-ae.tex
\begin{figure}
  \centering  
  \texttt{\zr1=x; y=1; || \zr2=y; x=\zr2;}
  \\[1ex]
  \begin{tikzpicture}[node distance=1em,baselinecenter]
    \zeventr{30}{init}{\init}{}
    \weventr{32}{Rx1}{\DR{x}{1}}{below left=of init}
    \xeventl{31}{Rx0}{\DR{x}{0}}{left of=Rx1,node distance=5em}
    \xeventl{35}{Wy0}{\DW{y}{1}}{below=of Rx0}
    \weventr{36}{Wy1}{\DW{y}{1}}{below=of Rx1}
    \xeventl{33}{Ry0}{\DR{y}{0}}{below right=of init}
    \xeventr{34}{Ry1}{\DR{y}{1}}{right of=Ry0,node distance=5em}
    \xeventl{37}{Wx0}{\DW{x}{0}}{below=of Ry0}
    \xeventr{38}{Wx1}{\DW{x}{1}}{below=of Ry1}
    \xpo{init}{Rx0}
    \xpo{init}{Rx1}
    \xpo{Rx0}{Wy0}
    \xpo{Rx1}{Wy1}
    \xpo{init}{Ry0}
    \xpo{init}{Ry1}
    \xpo{Ry0}{Wx0}
    \xpo{Ry1}{Wx1}
    \xconfl{Rx0}{Rx1}
    \xconfl{Ry0}{Ry1}
    \zrwj[out=180,in=35]{init}{Rx0}
    \zrwj[out=0,in=90]{init}{Ry0}
    \zrwj[out=340,in=200]{Wy0}{Ry1}
  \end{tikzpicture}
  \\[2ex]
  \begin{tikzpicture}[node distance=1em,baselinecenter]
    \zeventr{30}{init}{\init}{}
    \xeventr{32}{Rx1}{\DR{x}{1}}{below left=of init}
    \xeventl{31}{Rx0}{\DR{x}{0}}{left of=Rx1,node distance=5em}
    \xeventl{35}{Wy0}{\DW{y}{1}}{below=of Rx0}
    \xeventr{36}{Wy1}{\DW{y}{1}}{below=of Rx1}
    \yeventl{33}{Ry0}{\DR{y}{0}}{below right=of init}
    \zeventr{34}{Ry1}{\DR{y}{1}}{right of=Ry0,node distance=5em}
    \yeventl{37}{Wx0}{\DW{x}{0}}{below=of Ry0}
    \zeventr{38}{Wx1}{\DW{x}{1}}{below=of Ry1}
    \xpo{init}{Rx0}
    \xpo{init}{Rx1}
    \xpo{Rx0}{Wy0}
    \xpo{Rx1}{Wy1}
    \ypo{init}{Ry0}
    \zpo{init}{Ry1}
    \ypo{Ry0}{Wx0}
    \zpo{Ry1}{Wx1}
    \xconfl{Rx0}{Rx1}
    \yconfl{Ry0}{Ry1}
    \zrwj[out=180,in=35]{init}{Rx0}
    \zrwj[out=340,in=200]{Wy0}{Ry1}
    \zrwj{Wy1}{Ry1}
    \zrwj{Wx1}{Rx1}
  \end{tikzpicture}
  \\[2ex]
  \begin{tikzpicture}[node distance=1em,baselinecenter]
    \zeventr{30}{init}{\init}{}
    \zeventr{32}{Rx1}{\DR{x}{1}}{below left=of init}
    \yeventl{31}{Rx0}{\DR{x}{0}}{left of=Rx1,node distance=5em}
    \yeventl{35}{Wy0}{\DW{y}{1}}{below=of Rx0}
    \zeventr{36}{Wy1}{\DW{y}{1}}{below=of Rx1}
    \yeventl{33}{Ry0}{\DR{y}{0}}{below right=of init}
    \zeventr{34}{Ry1}{\DR{y}{1}}{right of=Ry0,node distance=5em}
    \yeventl{37}{Wx0}{\DW{x}{0}}{below=of Ry0}
    \zeventr{38}{Wx1}{\DW{x}{1}}{below=of Ry1}
    \ypo{init}{Rx0}
    \zpo{init}{Rx1}
    \ypo{Rx0}{Wy0}
    \zpo{Rx1}{Wy1}
    \ypo{init}{Ry0}
    \zpo{init}{Ry1}
    \ypo{Ry0}{Wx0}
    \zpo{Ry1}{Wx1}
    \yconfl{Rx0}{Rx1}
    \yconfl{Ry0}{Ry1}
    \xrwj{Wy1}{Ry1}
    \xrwj{Wx1}{Rx1}
  \end{tikzpicture}
  \caption{AE-order for \ref{eq:p3}}
  \label{fig:p3:ae}
\end{figure}


%% file: fig-sevcik532.tex
\begin{figure}
  \texttt{\normalsize{y=x; || if(y)\{x=y;\}\,else\,\{x=1;\}}}
  \\[-8ex]
        \begin{tikzpicture}[node distance=1em,baselinecenter]
          \zeventr{a}{init}{\init}{}
          \xeventl{b}{Rx0}{\DR{x}{0}}{left of=Rx1,node distance=5em}
          \xeventr{c}{Rx1}{\DR{x}{1}}{below left=of init}
          \xeventl{d}{aRy0}{\DR{y}{0}}{below right=of init}
          \weventr{e}{aRy1}{\DR{y}{1}}{right of=Ry0,node distance=8em}
          \xeventl{f}{Wy0}{\DW{y}{0}}{below=of Rx0}
          \xeventr{g}{Wy1}{\DW{y}{1}}{below=of Rx1}
          \xeventl{h}{aWx1}{\DW{x}{1}}{below=of aRy0}
          \weventl{i}{bRy0}{\DR{y}{0}}{below left=of aRy1}
          \weventr{j}{bRy1}{\DR{y}{1}}{below right=of aRy1}
          \weventl{k}{bWx0}{\DW{x}{0}}{below=of bRy0}
          \weventr{l}{bWx1}{\DW{x}{1}}{below=of bRy1}
          \xpo{init}{Rx0}
          \xpo{init}{Rx1}
          \xpo{Rx0}{Wy0}
          \xpo{Rx1}{Wy1}
          \xpo{init}{aRy0}
          \xpo{init}{aRy1}
          \xpo{aRy0}{aWx1}
          \xpo{aRy1}{bRy0}
          \xpo{aRy1}{bRy1}
          \xpo{bRy0}{bWx0}
          \xpo{bRy1}{bWx1}
          \xconfl{Rx0}{Rx1}
          \xconfl{aRy0}{aRy1}
          \xconfl{bRy0}{bRy1}
          \zrwj[out=180,in=35]{init}{Rx0}
          \zrwj[out=0,in=90]{init}{aRy0}
          \xrwj[out=30,in=70]{bWx1}{Rx1}
          \zrwj{aWx1}{Rx1}
          \xrwj[out=-45,in=-60]{Wy1}{aRy1}
          \xrwj[out=-45,in=-140]{Wy1}{bRy1}
        \end{tikzpicture}
  \\[-10ex]
        \begin{tikzpicture}[node distance=1em,baselinecenter]
          \zeventr{a}{init}{\init}{}
          \yeventl{b}{Rx0}{\DR{x}{0}}{left of=Rx1,node distance=5em}
          \zeventr{c}{Rx1}{\DR{x}{1}}{below left=of init}
          \xeventl{d}{aRy0}{\DR{y}{0}}{below right=of init}
          \xeventr{e}{aRy1}{\DR{y}{1}}{right of=Ry0,node distance=8em}
          \yeventl{f}{Wy0}{\DW{y}{0}}{below=of Rx0}
          \zeventr{g}{Wy1}{\DW{y}{1}}{below=of Rx1}
          \xeventl{h}{aWx1}{\DW{x}{1}}{below=of aRy0}
          \xeventl{i}{bRy0}{\DR{y}{0}}{below left=of aRy1}
          \xeventr{j}{bRy1}{\DR{y}{1}}{below right=of aRy1}
          \xeventl{k}{bWx0}{\DW{x}{0}}{below=of bRy0}
          \xeventr{l}{bWx1}{\DW{x}{1}}{below=of bRy1}
          \ypo{init}{Rx0}
          \zpo{init}{Rx1}
          \ypo{Rx0}{Wy0}
          \zpo{Rx1}{Wy1}
          \xpo{init}{aRy0}
          \xpo{init}{aRy1}
          \xpo{aRy0}{aWx1}
          \xpo{aRy1}{bRy0}
          \xpo{aRy1}{bRy1}
          \xpo{bRy0}{bWx0}
          \xpo{bRy1}{bWx1}
          \yconfl{Rx0}{Rx1}
          \xconfl{aRy0}{aRy1}
          \xconfl{bRy0}{bRy1}
          \zrwj[out=0,in=90]{init}{aRy0}
          \zrwj[out=0,in=90]{init}{bRy0}
          \xrwj[out=30,in=70]{bWx1}{Rx1}
          \xrwj{aWx1}{Rx1}
          \zrwj[out=-45,in=-60]{Wy1}{aRy1}
          \zrwj[out=-45,in=-140]{Wy1}{bRy1}
        \end{tikzpicture}
  \\[-10ex]
        \begin{tikzpicture}[node distance=1em,baselinecenter]
          \zeventr{a}{init}{\init}{}
          \yeventl{b}{Rx0}{\DR{x}{0}}{left of=Rx1,node distance=5em}
          \zeventr{c}{Rx1}{\DR{x}{1}}{below left=of init}
          \yeventl{d}{aRy0}{\DR{y}{0}}{below right=of init}
          \zeventr{e}{aRy1}{\DR{y}{1}}{right of=Ry0,node distance=8em}
          \yeventl{f}{Wy0}{\DW{y}{0}}{below=of Rx0}
          \zeventr{g}{Wy1}{\DW{y}{1}}{below=of Rx1}
          \yeventl{h}{aWx1}{\DW{x}{1}}{below=of aRy0}
          \xeventl{i}{bRy0}{\DR{y}{0}}{below left=of aRy1}
          \xeventr{j}{bRy1}{\DR{y}{1}}{below right=of aRy1}
          \xeventl{k}{bWx0}{\DW{x}{0}}{below=of bRy0}
          \xeventr{l}{bWx1}{\DW{x}{1}}{below=of bRy1}
          \ypo{init}{Rx0}
          \zpo{init}{Rx1}
          \ypo{Rx0}{Wy0}
          \zpo{Rx1}{Wy1}
          \ypo{init}{aRy0}
          \zpo{init}{aRy1}
          \ypo{aRy0}{aWx1}
          \xpo{aRy1}{bRy0}
          \xpo{aRy1}{bRy1}
          \xpo{bRy0}{bWx0}
          \xpo{bRy1}{bWx1}
          \yconfl{Rx0}{Rx1}
          \yconfl{aRy0}{aRy1}
          \xconfl{bRy0}{bRy1}
          \zrwj[out=0,in=90]{init}{bRy0}
          \xrwj[out=30,in=70]{bWx1}{Rx1}
          \xrwj[out=-45,in=-60]{Wy1}{aRy1}
          \zrwj[out=-45,in=-140]{Wy1}{bRy1}
        \end{tikzpicture}
  \\[-10ex]
        \begin{tikzpicture}[node distance=1em,baselinecenter]
          \zeventr{a}{init}{\init}{}
          \yeventl{b}{Rx0}{\DR{x}{0}}{left of=Rx1,node distance=5em}
          \zeventr{c}{Rx1}{\DR{x}{1}}{below left=of init}
          \yeventl{d}{aRy0}{\DR{y}{0}}{below right=of init}
          \zeventr{e}{aRy1}{\DR{y}{1}}{right of=Ry0,node distance=8em}
          \yeventl{f}{Wy0}{\DW{y}{0}}{below=of Rx0}
          \zeventr{g}{Wy1}{\DW{y}{1}}{below=of Rx1}
          \yeventl{h}{aWx1}{\DW{x}{1}}{below=of aRy0}
          \yeventl{i}{bRy0}{\DR{y}{0}}{below left=of aRy1}
          \zeventr{j}{bRy1}{\DR{y}{1}}{below right=of aRy1}
          \yeventl{k}{bWx0}{\DW{x}{0}}{below=of bRy0}
          \zeventr{l}{bWx1}{\DW{x}{1}}{below=of bRy1}
          \ypo{init}{Rx0}
          \zpo{init}{Rx1}
          \ypo{Rx0}{Wy0}
          \zpo{Rx1}{Wy1}
          \ypo{init}{aRy0}
          \zpo{init}{aRy1}
          \ypo{aRy0}{aWx1}
          \ypo{aRy1}{bRy0}
          \zpo{aRy1}{bRy1}
          \ypo{bRy0}{bWx0}
          \zpo{bRy1}{bWx1}
          \yconfl{Rx0}{Rx1}
          \yconfl{aRy0}{aRy1}
          \yconfl{bRy0}{bRy1}
          \xrwj[out=30,in=70]{bWx1}{Rx1}
          \xrwj[out=-45,in=-60]{Wy1}{aRy1}
          \xrwj[out=-45,in=-140]{Wy1}{bRy1}
        \end{tikzpicture}
  \caption{AE-order for {\v{S}ev\v{c}\'{\i}k \textsection{}5.3.2}}
  \label{fig:sevcik532}
\end{figure}
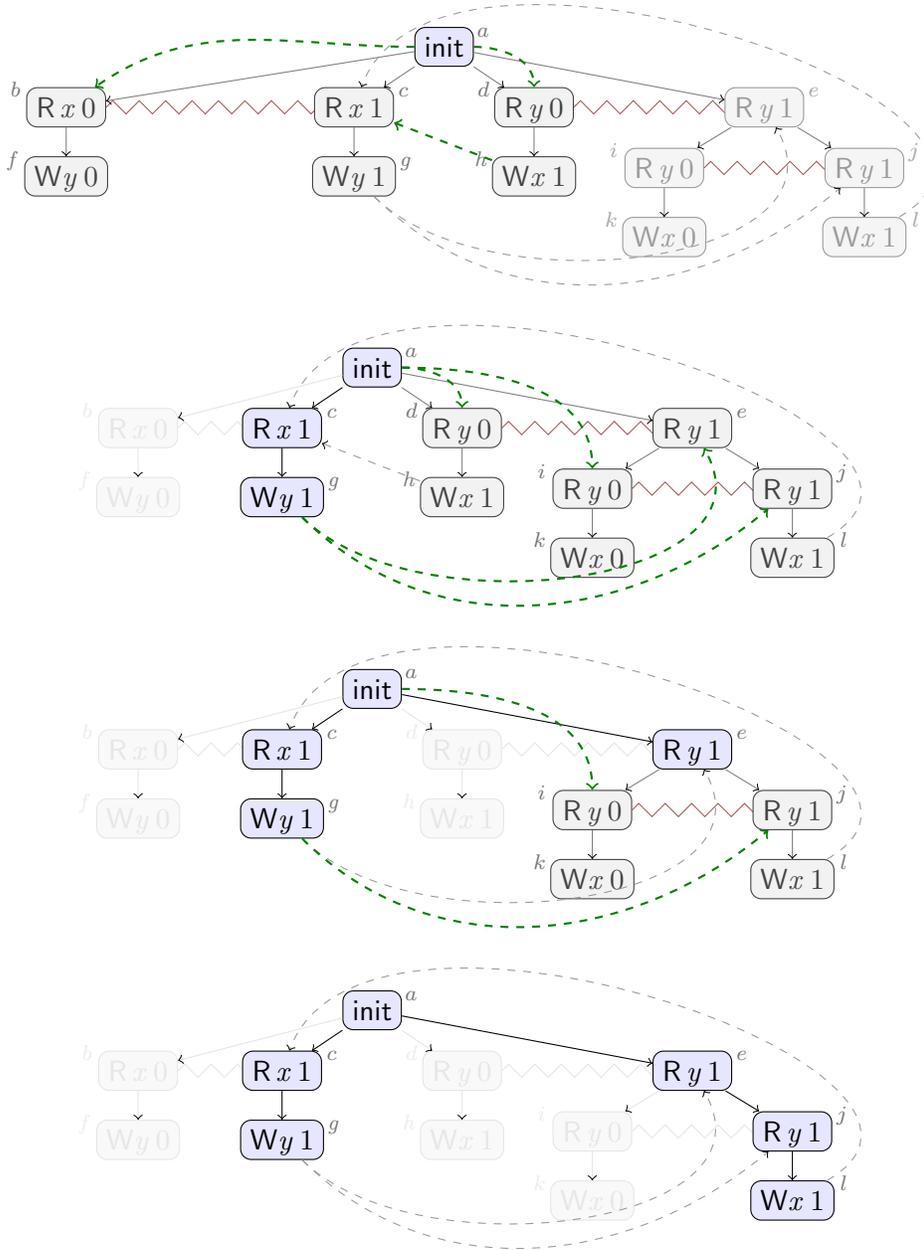


%% file: drf.tex
\section{Data-race-free event structures}
\label{sec:drf}


We say that $\ES'$ is an augmentation of $\ES$ if it has same events, conflict and
labels, and possibly more order.  Formally, $(E,{\le'},{\confl},\lambda)$ is
an augmentation of $(E,{\le},{\confl},\lambda)$ if ${\leq}\subseteq{\leq'}$.

It is straightforward to show that justification, acyclic justification and
well-justification are all reflected by augmentation.  For example, if $\ES'$
augments $\ES$ and $C$ is a well-justified configuration of $\ES'$ then $C$
is a well-justified configuration of $\ES$.

A \emph{sequential} memory event structure is one where,
for any events $d$ and $e$, either $d \le e$,
$e \le d$ or $d \confl e$.
A \emph{sequentially consistent} configuration of a memory event structure is a
justified configuration of a sequential augmentation of it.
That is, a configuration $C$ of $\ES$ is {sequentially consistent} if there
exists an augmentation $\ES'$ of $\ES$ such that $\ES'$ is sequential and $C$
is a justified configuration of $\ES'$.

Note that in a sequential memory event structure, if $d$ justifies $e$
then $d \le e$. It follows that any justified configuration of a
sequential memory event structure is well-justified, and hence that
any sequentially consistent configuration of a memory event structure
is well-justified.

The converse is not true.  There are well-justified configurations that are
not sequentially consistent, due to data races. For example, the
program \texttt{(w=1; || y=(w<=x); || z=(x<=w); || x=1;)} has semantics with
configuration:
\begin{small}
\begin{displaymath}
  \begin{tikzpicture}[node distance=1em]
    \zevent{init}{\init}{}
    \zevent{Rw1}{\DR{w}{1}}{below left=of init}
    \zevent{Rx1}{\DR{x}{1}}{below right=of init}
    \zevent{Rx0}{\DR{x}{0}}{below=of Rw1}
    \zevent{Rw0}{\DR{w}{0}}{below=of Rx1}
    \zevent{Ww1}{\DW{w}{1}}{left=2em of Rw1}
    \zevent{Wx1}{\DW{x}{1}}{right=2em of Rx1}
    \zevent{Wy0}{\DW{y}{0}}{below=of Rx0}
    \zevent{Wz0}{\DW{z}{0}}{below=of Rw0}
    \zpo{init}{Rw1}
    \zpo{init}{Ww1}
    \zpo{init}{Rx1}
    \zpo{init}{Wx1}
    \zpo{Rw1}{Rx0}
    \zpo{Rx1}{Rw0}
    \zpo{Rx0}{Wy0}
    \zpo{Rw0}{Wz0}
    \zrwj[out=255,in=30]{init}{Rx0}
    \zrwj[out=285,in=150]{init}{Rw0}
    \zrwj{Wx1}{Rx1}
    \zrwj{Ww1}{Rw1}
  \end{tikzpicture}
\end{displaymath}
\ignorespacesafterend\end{small}
This is acyclically justified, and hence well-justified, but not sequentially consistent.
In this section, we shall show that such data races
are the only source of configurations which are well-justified
but not sequentially consistent.

In a memory event structure, define concurrent events $d$ and $e$
to be a \emph{read-write race} whenever there is some $c \muconfleq e$
such that $d$ justifies $c$, as shown on the left below. 
%
Define concurrent events $d$ and $e$
to be a \emph{write-write race} whenever there is some $b \muconfleq c$
such that $d$ justifies $b$ and $e$ justifies $c$, as shown on the right below. 
\begin{align*}
  \begin{tikzpicture}[node distance=1em,baselinecenter]
    \zevent{d}{d}{}
    \zevent{c}{c}{below=of d}
    \zevent{e}{e}{right=2em of c}
    \zconfl{c}{e}
    \zrwj{d}{c}
  \end{tikzpicture}
  &&
  \begin{tikzpicture}[node distance=1em,baselinecenter]
    \zevent{d}{d}{}
    \zevent{e}{e}{right=2em of d}
    \zevent{b}{b}{below=of d}
    \zevent{c}{c}{below=of e}
    \zconfl{b}{c}
    \zrwj{d}{b}
    \zrwj{e}{c}
  \end{tikzpicture}
\end{align*}
Define a configuration to be \emph{data-race-free} when it contains no
read-write or write-write races.  Recall that $\muconfleq$ is reflexive;
thus, we may also have $c=e$ in the left diagram and $b=c$ in the right.

We state the theorem generally for any memory event structure that is
read-enabled and commutative.  The prototypical example, given in
\textsection\ref{sec:memory}, satisfies both these criteria.

\begin{wrapfigure}{r}{0.24\columnwidth}
  \centering
  \begin{tikzpicture}[node distance=1em,baselinecenter]
    \zevent{c}{c}{}
    \zevent{d}{d}{below left=of c}
    \zevent{e}{e}{below right=of c}
    \zpo{c}{e}
    \zconfl{d}{e}
    \zrwj{c}{d}
  \end{tikzpicture}
\end{wrapfigure}
A memory event structure is \emph{read-enabled} whenever, for any read
event $e$ there exists some $c \le e \muconfleq d$ such that $c$ justifies
$d$, as shown on the right. 
Any event structure that is the semantics of a program is read enabled.  
Read enabledness ensures that every read operation in a program can be satisfied by some
preceding write.  
This is true because of the $\init$
event which justifies reading $0$ on every variable and, therefore,
is in read-write conflict with every read.
For example, if $e$ is $\DR{x}{1}$, then it suffices to take $c$ to be
$\init{}$, since $\init{}$ justifies $\DR{x}{0}$, which is in primitive
conflict with $\DR{x}{1}$.

A memory event structure is \emph{commutative} whenever 
$c \muconfleq d$ and $d$ justifies $e$ 
implies
there exists $b \muconfleq e$
where $c$ justifies $b$, that is:
\begin{displaymath}
  \begin{tikzpicture}[node distance=1em,baselinecenter]
    \zevent{c}{c}{}
    \zevent{d}{d}{right=of c}
    \zevent{e}{e}{below=of d}
    \zconfl{c}{d}
    \zrwj{d}{e}
  \end{tikzpicture}
  \quad\text{implies}\quad
  \begin{tikzpicture}[node distance=1em,baselinecenter]
    \zevent{c}{c}{}
    \zevent{d}{d}{right=of c}
    \zevent{e}{e}{below=of d}
    \zevent{b}{b}{below=of c}
    \zconfl{c}{d}
    \zconfl{b}{e}
    \zrwj{d}{e}
    \zrwj{c}{b}
  \end{tikzpicture}
\end{displaymath}
If read and write actions are disjoint, then it follows immediately that 
any event structure that is the semantics of a program will be commutative,
since read is defined as a sum over all possible values.
Commutativity ensures that all writes can be read from.

Read-modify-write operators such as swap and fetch-and-add are commutative,
since these always write.
Compare-and-set ($\CAS$) is commutative if we interpret a failed $\CAS$ as both
read and write (of the old value), but not if we consider a failed $\CAS$ only
as a read.  For example, if failed $\CAS$ is considered a read, then
$(\CAS\,{x}\,{0}\,{1})$ generates the following event structure for bit register
$x$, where $\DRMW{x}{0}{1}$ denotes a successful $\CAS$ and $\DR{x}{1}$
denotes a failed $\CAS$.
\begin{gather*}
        \begin{tikzpicture}[node distance=1em,baselinecenter]
          \zevent{init}{\init}{}
          \zevent{x0}{\DRMW{x}{0}{1}}{below left=of init}
          \zevent{x1}{\DR{x}{1}}{below right = of init}
          \zpo{init}{x0}
          \zpo{init}{x1}
          \zconfl{x0}{x1}
        \end{tikzpicture}  
\end{gather*}
The $\DRMW{x}{0}{1}$ event may justify some other read of $x$; however, the
minimal conflicting event $\DR{x}{1}$ is a plain read, which justifies nothing.
%
%
We leave weakening commutativity as future work.

\begin{theorem}[DRF]\ 
  \label{thm:drf}
  In any commutative read-enabled memory event structure, 
  if all sequentially consistent configurations are data-race-free,
  then all well-justified configurations are sequentially consistent.
\begin{proof}
  Define a configuration $C$ to be \emph{pre-justified} if every read
  action $e\in C$ is justified by a write action $d \in C$ where $d \le e$.
  It is routine to show that any pre-justified configuration is sequentially
  consistent.
  The core lemma of the proof follows \citep{DBLP:journals/toplas/Lochbihler13},
  which is that if $C$ is pre-justified, and $C$ justifies $D$, then $D$ is pre-justified.
  After this, the proof is routine: we first show that if $C$ is pre-justified and $C \acext D$
  then $D$ is pre-justified; then that if $C$ is pre-justified and $C \aeext D$ then 
  $D$ is pre-justified. The result follows, since $\emptyset$ is trivially pre-justified.  
  This proof has been mechanized in Agda.
\end{proof}
\end{theorem}


%% file: invariants.tex
\section{Invariants}
\label{sec:invariants}

While there is no formal definition of ``thin-air read''
\citep{DBLP:conf/esop/BattyMNPS15}, the examples point to a failure of
inductive reasoning, typically due to a cycle in the union of the causal and
data dependency orders.  In order to establish that these forms of thin-air
read are impossible, it is sufficient to show that is possible to reason
inductively.  In this section, we show that well-justification enables
inductive reasoning.

We consider a limited form of invariant reasoning, which is strong enough to
capture non-temporal safety properties, such as type safety.  Given a suitable
notion of formula, $\phi$, we show that if, in every 
configuration of $\ES$, the \emph{read} events satisfy $\phi$, then, in every
configuration of $\ES$, \emph{all} events satisfy $\phi$.
Significantly, the result can be applied without reasoning about well-justification.

To keep the setting as simple as possible, we consider logics over {labels}
rather than events.  In order to establish the result, we must restrict
attention to logics that are subset closed.  This allows the expression of
certain safety properties such as $x\mathord{\neq} 1$, but not liveness properties
such as $x\mathord=1$.
For a label set $\Sigma$, 
a \emph{program logic} $(\Phi,{\vDash})$ consists of:
\begin{itemize}
\item a set $\Phi$ (the \emph{formulae}), and
\item a binary relation $\vDash$ between $\mathcal{P}(\Sigma)$ and $\Phi$
  (\emph{satisfaction}).
\end{itemize}
A formula $\phi$ is \emph{subset closed} whenever
$A \subseteq B \vDash \phi$ implies $A \vDash \phi$. It is
\emph{satisfiable} whenever $A \vDash \phi$ for some $A$.
It \emph{respects justification} whenever $A$ justifies $B$ and
$A \vDash \phi$ implies $B \vDash \phi$.

For any configuration $C$, let $\Sigma(C)$ be the labels of $C$:
\begin{displaymath}
  \Sigma(C) = \{ \lambda(e) \mid e \in C \}
\end{displaymath}
A formula $\phi$ is an \emph{invariant} of a memory event structure
whenever $\Sigma(C) \cap R \vDash \phi$ implies $\Sigma(C) \vDash \phi$ for any
configuration $C$ (recalling that $R$ is the set of all read actions).

A formula $\phi$ is a \emph{tautology} of a memory event structure
whenever $\Sigma(C) \vDash \phi$ for any well-justified
configuration $C$.

\begin{theorem}
  \label{thm:inv}
  For any satisfiable, subset-closed $\phi$ which respects justification,
  if $\phi$ is an invariant of $\ES$
  then $\phi$ is a tautology of $\ES$.
\begin{proof}
  Mechanized in Agda.
\end{proof}
\end{theorem}

In the remainder of this section, we consider an example logic.  Let $\Types$
be a set of \emph{type names}, ranged over by $\atyp$.  The set $\Phi$ of
formulae is generated by the following BNF.
\begin{displaymath}
  \phi\C\psi \BNFDEF
  \pneq{x}{v} \BNFSEP 
  \ptype{x}{\atyp} \BNFSEP 
  \ptrue \BNFSEP \pfalse \BNFSEP \phi \pand \psi \BNFSEP \phi \por \psi
\end{displaymath}
Given a semantics $V_\tau\subseteq V$ for each type $\tau$,
let $\vDash$ be the obvious
satisfaction relation generated by the following rules for the atoms.
\begin{itemize}
\item $A \vDash\pneq{x}{v}$ when for any $a \in A$,
  if $a=\DR{x}{w}$ or $a=\DW{x}{w}$,
  then $w \neq v$.
\item $A \vDash\ptype{x}{\atyp}$ when
  for any $a\in A$, if $a=\DR{x}{v}$ or $a=\DW{x}{v}$,
  then $v \in V_\tau$.
\end{itemize}
Note that the logic is satisfiable and subset closed and thus satisfies the
criteria of Theorem~\ref{thm:inv}.

Suppose that we attempt to show that $\phi_1=(\pneq{x}{1}\pand\pneq{y}{1})$ is a
tautology for $\text{\ref{eq:p1}}$.  Recall the event
structure for this program, given in the introduction.
\begin{gather*}
  \texttt{(y=x; || x=y;)}
  \\
  \begin{tikzpicture}[node distance=1em,baselinecenter]
    \zeventr{10}{init}{\init}{}
    \zeventr{12}{Rx1}{\DR{x}{1}}{below left=of init}
    \zeventl{11}{Rx0}{\DR{x}{0}}{left of=Rx1,node distance=5em}
    \zeventl{15}{Wy0}{\DW{y}{0}}{below=of Rx0}
    \zeventr{16}{Wy1}{\DW{y}{1}}{below=of Rx1}
    \zeventl{13}{Ry0}{\DR{y}{0}}{below right=of init}
    \zeventr{14}{Ry1}{\DR{y}{1}}{right of=Ry0,node distance=5em}
    \zeventl{17}{Wx0}{\DW{x}{0}}{below=of Ry0}
    \zeventr{18}{Wx1}{\DW{x}{1}}{below=of Ry1}
    \zpo{init}{Rx0}
    \zpo{init}{Rx1}
    \zpo{Rx0}{Wy0}
    \zpo{Rx1}{Wy1}
    \zpo{init}{Ry0}
    \zpo{init}{Ry1}
    \zpo{Ry0}{Wx0}
    \zpo{Ry1}{Wx1}
    \zconfl{Rx0}{Rx1}
    \zconfl{Ry0}{Ry1}
  \end{tikzpicture}
\end{gather*}
Note that any configuration which includes write events \zz{16} or \zz{18},
must also include read events \zz{12} or \zz{14}. Thus, if the read events
satisfy $\phi_1$ then the write events satisfy $\phi_1$, and so 
$\phi_1$ is invariant for \ref{eq:p1}. Thus, by
Theorem~\ref{thm:inv}, $\phi_1$ is a tautology for \ref{eq:p1}. 

For \text{\ref{eq:p3}}, instead, $\phi_1$
fails.
Recall the event structure for this program, also given in the introduction.
\begin{gather*}
  \texttt{(r=x; y=1; || x=y;)}
  \\
  \begin{tikzpicture}[node distance=1em,baselinecenter]
    \zeventr{30}{init}{\init}{}
    \zeventr{32}{Rx1}{\DR{x}{1}}{below left=of init}
    \zeventl{31}{Rx0}{\DR{x}{0}}{left of=Rx1,node distance=5em}
    \zeventl{35}{Wy0}{\DW{y}{1}}{below=of Rx0}
    \zeventr{36}{Wy1}{\DW{y}{1}}{below=of Rx1}
    \zeventl{33}{Ry0}{\DR{y}{0}}{below right=of init}
    \zeventr{34}{Ry1}{\DR{y}{1}}{right of=Ry0,node distance=5em}
    \zeventl{37}{Wx0}{\DW{x}{0}}{below=of Ry0}
    \zeventr{38}{Wx1}{\DW{x}{1}}{below=of Ry1}
    \zpo{init}{Rx0}
    \zpo{init}{Rx1}
    \zpo{Rx0}{Wy0}
    \zpo{Rx1}{Wy1}
    \zpo{init}{Ry0}
    \zpo{init}{Ry1}
    \zpo{Ry0}{Wx0}
    \zpo{Ry1}{Wx1}
    \zconfl{Rx0}{Rx1}
    \zconfl{Ry0}{Ry1}
  \end{tikzpicture}
\end{gather*}
The configuration $\set{\zz{30}\C\zz{31}\C\zz{35}}$ fails to satisfy
$\phi_1$ even though its only read event $\zz{31}$ satisfies $\phi_1$.

These examples can be adapted to show reasoning using types.  Let $0$ be the
unique value of type $\Unit$.  Then \ref{eq:p1} satisfies the typing
$\ptype{x}{\Unit}\pand\ptype{y}{\Unit}$, but \ref{eq:p2} does not.

%% file: fencing.tex
\section{Fencing}
\label{sec:fencing}

In \textsection\ref{sec:memory}, we noted that a memory alphabet includes
\emph{synchronized justification}, such as lock release and acquire; however,
we have not made any use of synchronization up to now. In this section, we
develop a notion of fencing, in which synchronized justifications
contribute to causal order.

We model lock-based synchronization in a very simple setting.  We assume
that there is only one, statically allocated lock and that lock release
always occurs in the same thread as the previous acquire.  The latter
assumption ensures that each release causally follows the corresponding acquire.

The memory alphabet from \textsection\ref{sec:memory} is modified to include
acquire and release actions, which are considered both read and write
actions.  (We comment on this design in \textsection\ref{sec:open}.)
\begin{alignat*}{2}
  \Sigma & = R \cup W \\
  R & = \{ \DR{x}{v}, \LA{v}, \LR{v} \mid x \in X, v \in V \} \\
  W & = \{ \DW{x}{v}, \LA{v}, \LR{v} \mid x \in X, v \in V \} \cup \{ \init \}
\end{alignat*}
The semantics of locking actions are as follows.
\begin{alignat*}{2}
  \T{ \pacq{M} T } ρ &\eqdef \esprefix{\LA{\M{M}{ρ}}}{\T{T}{ρ}} \\
  \T{ \prel{M} T } ρ &\eqdef \esprefix{\LR{\M{M}{ρ}}}{\T{T}{ρ}} 
\end{alignat*}
The justification relation, $J$, now includes lock actions:
\begin{alignat*}{2}
  J & =   \{ (\init,\DR{x}{0}), (\DW{x}{v},\DR{x}{v}) \mid x \in X, v \in V \}\\
    &\quad \cup \{ (\init, \LA{v}), (\LA{v},\LR{v}), (\LR{v},\LA{v}) \}
\end{alignat*}
The synchronized justification relation, $K$, is restricted to lock actions.
\begin{alignat*}{2}
  K & =   
    \{ (\init, \LA{v}), (\LA{v},\LR{v}), (\LR{v},\LA{v}) \}
\end{alignat*}


Recall from \textsection\ref{sec:memory} that event $d$ \emph{justifies}
event $e$ whenever
\begin{enumerate*}
\item $(\lambda(d),\lambda(e)) \in J$,
\item $d$ does not follow $e$,
\item $d$ is not in conflict with $e$, and
\item there is no intervening $b$ between $d$ and $e$ that justifies an event
  in primitive conflict with $e$.
\end{enumerate*}

We say event $d$ \emph{synchronously justifies} event $e$ whenever $d$
justifies $e$ and $(\lambda(d),\lambda(e)) \in K$.

A \emph{fenced} memory event structure is one where,
for any events $d$ and $e$,
if $d$ synchronously justifies $e$ then $d \le e$.

A \emph{well-fenced} configuration of a memory event
structure is a well-justified configuration of a 
fenced augmentation of it.
That is, a configuration $C$ of $\ES$ is well-fenced if there
exists a augmentation $\ES'$ of $\ES$ such that $\ES'$ is fenced and $C$
is a justified configuration of $\ES'$. 

To show that a configuration is well-fenced, the player must first augment the event structure
so that it is fenced.
Then the inductive argument for well-justification proceeds as before.

For example, consider the following program, \ref{eq:p5}.
\begin{scope}
\begin{gather*}
  \tag{\ensuremath{P_5}} \label{eq:p5}
  \texttt{\pacq{2} x=1; x=0; \prel{2} || \pacq{2} r=x; \prel{2}}
  \\[1ex]
  \begin{tikzpicture}[node distance=1em,baselinecenter]
    \zeventr{50}{init}{\init}{}
    \zeventl{51}{A0}{\LA{2}}{below left=of init,xshift=-3mm}
    \zeventl{53}{Wx1}{\DW{x}{1}}{below=of A0}
    \zeventl{56}{Wx0}{\DW{x}{0}}{below=of Wx1}
    \zeventl{57}{R0}{\LR{2}}{below=of Wx0}
    \zeventr{52}{A1}{\LA{2}}{below right=of init,xshift=3mm}
    \zeventl{54}{Rx0}{\DR{x}{0}}{below left=of A1,xshift=2mm,yshift=-5mm}
    \zeventr{55}{Rx1}{\DR{x}{1}}{below right=of A1,xshift=-2mm,yshift=-5mm}
    \zeventl{58}{R10}{\LR{2}}{below=of Rx0,yshift=-4.5mm}
    \zeventr{59}{R11}{\LR{2}}{below=of Rx1,yshift=-4.5mm}
    \zpo{init}{A0}
    \zpo{A0}{Wx1}
    \zpo{Wx1}{Wx0}
    \zpo{Wx0}{R0}
    \zpo{init}{A1}
    \zpo{A1}{Rx0}
    \zpo{A1}{Rx1}
    \zconfl{Rx0}{Rx1}
    \zpo{Rx0}{R10}
    \zpo{Rx1}{R11}
  \end{tikzpicture}
\end{gather*}
\end{scope}
Whereas the second thread can read $1$ in a well-justified configuration,
this is not possible in a well-fenced configuration. 
There are two possible fencings.  One includes the augmentation
$\zz{57}\le\zz{52}$, and the other includes $\zz{58}\le\zz{51}$ and $\zz{59}\le\zz{51}$.
In either case, $\zz{54}$ can be justified, but $\zz{55}$ cannot.  Therefore,
there is no well-fenced configuration that includes $\zz{55}$.  

Note that any sequential memory event structure is fenced.  It follows
that any sequential augmentation of a memory event structure is a fenced augmentation of
it, and hence that any sequentially consistent configuration
is a well-fenced configuration.

Now, if every justification is synchronized (that is
$J=K$) we have that every well-fenced configuration
is sequentially consistent, but in general this is not true.
In particular, if there is no synchronization
(that is $K=\emptyset$) then every well-justified configuration
is well-fenced.

Fortunately, the proof of the DRF Theorem for well-fenced
configurations follows directly from the DRF theorem for well-justified
configurations: we just use DRF on each fencing.

\begin{theorem}[DRF]
  In any commutative read-enabled memory event structure, 
  if all sequentially consistent configurations are data-race-free,
  then all well-fenced configurations are sequentially consistent.
\begin{proof}
  Follows directly from Theorem~\ref{thm:drf}.
\end{proof}
\end{theorem}


%% file: outro.tex
\section{Limitations}
\label{sec:outro}



As we noted in \S\ref{sec:memory}, one possible goal for a memory model is
that an execution should be allowed if there exists an augmentation, such as
thread inlining, that allows it.  A related goal is that an execution should
be allowed if it is allowed for a subset of threads.  These are non-goals for
our semantics, which, indeed, fails to validate them.

Both adding new threads and reducing causal order afford greater flexibility
in choosing configurations.  The additional flexibility can be exploited by
the player, validating additional executions.  But it can also be exploited by
the opponent, eliminating executions that would be possible without the
additional flexibility.  Consider a variant of the program given in
Example~\ref{ex:p3}, which is similar to TC9 from \citep{PughWebsite}:
\begin{gather*}
    \label{eq:tc9}\tag{TC9$'$}
  \texttt{\zr1=x; if(\zr1<2)\{y=1;\} || x=2; || \zr2=y; x=\zr2;}
  \\
  \begin{tikzpicture}[node distance=1em,baselinecenter]
    \zevent{init}{\init}{}
    \zevent{Wx2a}{\DW{x}{2}}{below=of init}
    \zevent{Rx2}{\DR{x}{2}}{left=2em of Wx2a}
    \zevent{Rx1}{\DR{x}{1}}{left of=Rx2,node distance=5em}
    \zevent{Rx0}{\DR{x}{0}}{left of=Rx1,node distance=5em}
    \zevent{Wy0}{\DW{y}{1}}{below=of Rx0}
    \zevent{Wy1}{\DW{y}{1}}{below=of Rx1}
    \zevent{Ry0}{\DR{y}{0}}{right=2em of Wx2a}
    \zevent{Ry1}{\DR{y}{1}}{right of=Ry0,node distance=5em}
    \zevent{Ry2}{\DR{y}{2}}{right of=Ry1,node distance=5em}
    \zevent{Wx0}{\DW{x}{0}}{below=of Ry0}
    \zevent{Wx1}{\DW{x}{1}}{below=of Ry1}
    \zevent{Wx2}{\DW{x}{2}}{below=of Ry2}
    \zpo[out=180,in=20]{init}{Rx0}
    \zpo[out=185,in=20]{init}{Rx1}
    \zpo{init}{Rx2}
    \zpo{Rx0}{Wy0}
    \zpo{Rx1}{Wy1}
    \zpo{Ry2}{Wx2}
    \zpo{init}{Ry0}
    \zpo[out=-10,in=160]{init}{Ry1}
    \zpo[out=-5,in=160]{init}{Ry2}    
    \zpo{init}{Wx2a}
    \zpo{Ry0}{Wx0}
    \zpo{Ry1}{Wx1}
    \zconfl{Rx0}{Rx1}
    \zconfl{Rx1}{Rx2}
    \zconfl{Ry0}{Ry1}
    \zconfl{Ry1}{Ry2}
  \end{tikzpicture}
\end{gather*}
Relative to Example~\ref{ex:p3}, we have added a thread that writes $2$ to
$x$.  To forbid configurations that include $\DR{y}{1}$, it is sufficient for
the opponent to chose a configuration that includes $\DW{x}{2}$ and
$\DR{x}{2}$; clearly such a configuration can be acyclically justified.  This
opponent move is unavailable, however, if the new thread is removed or if it
is inlined after either of the other threads; in this case, $\DR{y}{1}$ is
possible, following the reasoning of Example~\ref{ex:p3}.

Of the remaining nineteen test cases from \citep{PughWebsite}, our semantics
disagrees with TC3, TC7 and TC11.  In all cases, the reason is the same.  We
discuss TC7, which best elucidates the issues.
\begin{small}\begin{gather*}
    \label{eq:tc7}\tag{\normalsize{}TC7}
  \texttt{\normalsize{r=z; y=x; || z=y; x=1;}}
  \\
  \begin{tikzpicture}[node distance=1em,baselinecenter]
    \zeventr{a}{init}{\init}{}
    \zeventr{c}{Rz1}{\DR{z}{1}}{below left=of init,xshift=-3mm}
    \zeventl{b}{Rz0}{\DR{z}{0}}{left of=Rz1,node distance=9em}
    \zeventr{g}{0Rx1}{\DR{x}{1}}{below right=of Rz0,xshift=-4mm}
    \zeventl{f}{0Rx0}{\DR{x}{0}}{below left=of Rz0,xshift=4mm}
    \zeventl{l}{0Wy0}{\DW{y}{0}}{below=of 0Rx0}
    \zeventr{m}{0Wy1}{\DW{y}{1}}{below=of 0Rx1}
    \zeventr{i}{1Rx1}{\DR{x}{1}}{below right=of Rz1,xshift=-4mm}
    \zeventl{h}{1Rx0}{\DR{x}{0}}{below left=of Rz1,xshift=4mm}
    \zeventl{n}{1Wy0}{\DW{y}{0}}{below=of 1Rx0}
    \zeventr{o}{1Wy1}{\DW{y}{1}}{below=of 1Rx1}
    \zeventl{d}{Ry0}{\DR{y}{0}}{below right=of init}
    \zeventr{e}{Ry1}{\DR{y}{1}}{right of=Ry0,node distance=4em}
    \zeventl{j}{Wz0}{\DW{z}{0}}{below=of Ry0}
    \zeventr{k}{Wz1}{\DW{z}{1}}{below=of Ry1}
    \zeventl{p}{Wx0}{\DW{x}{1}}{below=of Wz0}
    \zeventr{q}{Wx1}{\DW{x}{1}}{below=of Wz1}
    \zpo{init}{Rz0}
    \zpo{init}{Rz1}
    \zconfl{0Rx0}{0Rx1}
    \zpo{Rz0}{0Rx0}
    \zpo{Rz0}{0Rx1}
    \zpo{0Rx0}{0Wy0}
    \zpo{0Rx1}{0Wy1}
    \zconfl{1Rx0}{1Rx1}
    \zpo{Rz1}{1Rx0}
    \zpo{Rz1}{1Rx1}
    \zpo{1Rx0}{1Wy0}
    \zpo{1Rx1}{1Wy1}
    \zpo{init}{Ry0}
    \zpo{init}{Ry1}
    \zpo{Ry0}{Wz0}
    \zpo{Ry1}{Wz1}
    \zpo{Wz0}{Wx0}
    \zpo{Wz1}{Wx1}
    \zconfl{Rz0}{Rz1}
    \zconfl{Ry0}{Ry1}
  \end{tikzpicture}
\end{gather*}
\ignorespacesafterend
\end{small}
The question is whether all of the reads can be resolved to $1$.  This fails
under our semantics: there is no well-justified configuration that includes
events \zz{c}, \zz{e} and \zz{i}, all of which read $1$.  In order to well-justify
such a configuration, one must first resolve the conflict on
$x$, then $y$ and finally $z$.   But this strategy fails immediately after
resolving $x$.  

Starting from the empty configuration, all acyclically justified
configurations can be extended to include either \zz{p} or \zz{q}, and thus
the player can select a configuration that includes either \zz{g} or \zz{i}.
Suppose the player selects \zz{i}.  Since configurations are downclosed, the
player must also select \zz{c}; however \zz{c} is not acyclically justified when
the opponent selects \zz{p}.  Symmetrically, \zz{b} is not acyclically justified when
the opponent selects \zz{q}.  Thus the player cannot resolve the conflict on $y$.

The failure of TC7 indicates a failure to validate the reordering of
independent reads.  To see this, consider the program in which the first
thread is rewritten.
\begin{small}
  \begin{gather*}
  \texttt{\normalsize{}y=x; r=z; || z=y; x=1;}
  \\
  \begin{tikzpicture}[node distance=1em,baselinecenter]
    \zeventr{a'}{init}{\init}{}
    \zeventr{c'}{Rz1}{\DR{x}{1}}{below left=of init,xshift=-3mm}
    \zeventl{b'}{Rz0}{\DR{x}{0}}{left of=Rz1,node distance=9em}
    \zeventr{h'}{Wy1}{\DW{y}{1}}{below=of Rz1}
    \zeventl{f'}{Wy0}{\DW{y}{0}}{below=of Rz0}
    \zeventr{m'}{0Rx1}{\DR{z}{1}}{below right=of Wy0,xshift=-4mm}
    \zeventl{l'}{0Rx0}{\DR{z}{0}}{below left=of Wy0,xshift=4mm}
    \zeventr{o'}{1Rx1}{\DR{z}{1}}{below right=of Wy1,xshift=-4mm}
    \zeventl{n'}{1Rx0}{\DR{z}{0}}{below left=of Wy1,xshift=4mm}
    \zeventl{d'}{Ry0}{\DR{y}{0}}{below right=of init}
    \zeventr{e'}{Ry1}{\DR{y}{1}}{right of=Ry0,node distance=4em}
    \zeventl{j'}{Wz0}{\DW{z}{0}}{below=of Ry0}
    \zeventr{k'}{Wz1}{\DW{z}{1}}{below=of Ry1}
    \zeventl{p'}{Wx0}{\DW{x}{1}}{below=of Wz0}
    \zeventr{q'}{Wx1}{\DW{x}{1}}{below=of Wz1}
    \zpo{init}{Rz0}
    \zpo{init}{Rz1}
    \zconfl{0Rx0}{0Rx1}
    \zpo{Wy0}{0Rx0}
    \zpo{Wy0}{0Rx1}
    \zpo{Rz0}{Wy0}
    \zconfl{1Rx0}{1Rx1}
    \zpo{Wy1}{1Rx0}
    \zpo{Wy1}{1Rx1}
    \zpo{Rz1}{Wy1}
    \zpo{init}{Ry0}
    \zpo{init}{Ry1}
    \zpo{Ry0}{Wz0}
    \zpo{Ry1}{Wz1}
    \zpo{Wz0}{Wx0}
    \zpo{Wz1}{Wx1}
    \zconfl{Rz0}{Rz1}
    \zconfl{Ry0}{Ry1}
  \end{tikzpicture}
\end{gather*}
\ignorespacesafterend
\end{small}
In this case, the player can choose \zz{c'}, then \zz{e'}, then \zz{o'}, as
required.

We now sketch a proposal to address this issue.
On sets of events, define $C$ is \emph{compatible} with $D$ whenever there is no
$c\in C$ and $d\in D$ such that $c \muconfl d$.
Define $C$ is \emph{consistent} whenever $C$ is compatible with $C$.

\begin{nproposal}
  Modify Definitions~\ref{def:acyclic} and \ref{def:well} to range over
  consistent sets, rather than configurations.
 
  A configuration $C$ is \emph{alt-well-justified} whenever $C$ is justified and
  there exists a consistent set $D$ such that $\emptyset \aeext D \supseteq C$.
\end{nproposal}

By this proposal, the definition of {alt-\AEjustifies{}} is as follows: On
consistent sets, $C$ \emph{alt-\AEjustifies{}} $D$ whenever for any
$C \acext C'$ there exists a $C' \acext C''$ such that $C''$ justifies $D$.

The configuration of TC7 that always reads $1$ is alt-well-justified. Starting
from the empty set,
\begin{math}
  C_1=\set{\zz{a},\zz{g},\zz{i}}
\end{math}
is alt-\AEjustified{}, since every consistent set that extends $\emptyset$ (via $\acext$) can be further extended
to include either \zz{p} or \zz{q}, both labeled $\DW{x}{1}$.  
Although \zz{g} and \zz{i} are in conflict, they are not in primitive
conflict, and therefore may both be included in a consistent set---this is
the key difference between well- and alt-well-justification.
From $C_1$, 
\begin{math}
  C_2=\set{\zz{a},\zz{g},\zz{i}, \zz{e}}
\end{math}
is alt-\AEjustified{}, since we can always extend
to include either \zz{m} or \zz{o}, both labeled $\DW{y}{1}$.  
From $C_2$, 
\begin{math}
  C_3=\set{\zz{a},\zz{g},\zz{i}, \zz{e}, \zz{c}}
\end{math}
is alt-\AEjustified{}, since we can always extend
to include \zz{k}, labeled $\DW{z}{1}$.  
Thus $\emptyset \aeext C_3$ and therefore, since writes do not require justification,
\begin{math}
  \emptyset \aeext 
  \set{\zz{a},\zz{g}, \zz{m}, \zz{i}, \zz{o}, \zz{e}, \zz{k}, \zz{q}, \zz{c}}
  \supseteq
  \set{\zz{a},\zz{i}, \zz{o}, \zz{e}, \zz{k}, \zz{q}, \zz{c}},
\end{math}
as required.




By hand calculation, alt-well-justification agrees with all nineteen test
cases; therefore, the definition looks quite promising.  The question of
whether DRF holds for alt-well-justification remains open.  The inductive
structure of the proof of DRF relies on the fact that configurations are
down closed.  Since consistent sets are not necessarily down closed, the proof
strategy fails.

\section{Related work}
\label{sec:related}

Our model is inspired by the Java Memory Model, first described by
\citet{DBLP:conf/popl/MansonPA05} and explored by \citet{SevcikThesis}.
\citet{DBLP:journals/toplas/Lochbihler13} provides mechanized proofs, in
addition to an encyclopedic survey and history.

The use of universal quantification over configurations in the definition of
\AEjustification{} is novel in this work. Prior definitions for Java-like
languages are purely existential in their quantification over possible
executions
\citep{DBLP:conf/popl/MansonPA05,DBLP:conf/esop/CenciarelliKS07,2010-gosrmm}.
This leads to differences between our model the JMM, such as the 
execution of Example~\ref{ex:sevcik532}, which is allowed in our model, but
disallowed by the JMM \citep[\textsection{}5.3.2]{SevcikThesis}.

\citet{DBLP:conf/esop/BattyMNPS15} describe the problem of thin-air
executions and provide a detailed review of the literature.  They show the
limitations of syntactic notions of dependency, as defined, for example, in
C++ \citep{DBLP:conf/pldi/BoehmA08,DBLP:conf/popl/BattyOSSW11}.  In
particular, for C++ they show that there is no per-candidate-execution
condition that precludes thin-air reads while allowing sufficient
optimization.  This problem is avoided in hardware models, such as
\citep{AlglaveThesis,AlglaveMT14}, which are not subject to compiler
optimization and therefore can encode the specific notion of dependency
provided by the hardware at hand.

Event structures have appeared in other attempts to formalize relaxed
memory semantics: 
\begin{itemize}
\item Our first attempt to model relaxed memory using labeled prime event
  structures was presented at a workshop in Cambridge \citep{talk14}.  Our
  proposal was to allow configurations that are \emph{relaxed justified}:
  Define a \emph{pre-configuration} to be a $\le$-downclosed set of events.
  A pre-configuration is \emph{relaxed justified} when it has a total order
  such that any non-initial event has a justifier that precedes it in the total
  order.  A justified configuration is \emph{relaxed justified} when it is
  included in a relaxed justified pre-configuration.  We abandoned this
  approach after \citet{rng} discovered that the program \texttt{(y=x+1; ||
    x=y;)} may generate any integer under this model.  This example is since
  referred to as the Random Number Generator (RNG).

\item
\cite{Sewell16} defined a semantics for C/C++ relaxed atomics using event
structures to model the state of a single thread, with the goal of providing
an semantic model of single-threaded optimization.  \emph{Deordering} allows
removal causal order between events of a thread that are not causally
related; for example, replacing \texttt{r=x;y=1} with \texttt{r=x || y=1}.
\emph{Merging} allows collapsing redundant reads into a single event; for
example, replacing \texttt{r1=x;r2=x;$C$} with \texttt{r1=x;$C'$}, where $C'$
is derived from $C$ by replacing occurrences of \texttt{r2} by \texttt{r1}.
\citet{Kang:2017:PSR:3009837.3009850} observe that this model is not robust
with respect to all of the relaxations permitted under hardware models.
\item
\citet{Castellan} presented a semantics for relaxed memory using event
structures and a restricted form of interleaving.  In this work, the
\emph{open} semantics of processes is as one expects for an event structure,
with parallel processes unordered by causality.  In the \emph{closed}
semantics, all events on the same memory cell are related by either conflict
or causality.  Although interleaving is only required for events on the same
memory cell, the process of computing the semantics may cause an exponential
increase in the state space.
\end{itemize}
These works can be classified by the role of universal quantification:
\citeauthor{Sewell16} use universal quantification to define
deordering. \citeauthor{Castellan} uses it to define interleaving.  We define
\AEjustification{} using universal quantification over the possible moves of
the opponent.

Unique among these work, we use the standard denotational semantics of
parallel composition for event structures.

Since the initial publication of this paper,
\citet{DBLP:conf/cgo/ChakrabortyV17} defined an operational semantics for a
subset of LLVM, which they extended in \citeyear{CV19}.  This semantics uses
``event structures'' to record potentially conflicting operational steps ---
these are close in  spirit to the pre-configurations of
\citep{talk14}.  As a result, this model is susceptible to the bad RNG
execution discussed above.  To avoid it, \citeauthor{CV19} restrict attention
to pre-configurations where are all events are \emph{visible}: event $e$ is
visible if for every write that precedes $e$ in causal+justification order,
there is a write with the same label that does not conflict with $e$.

\citet{pride} showed how to simulate several memory models using fragments of
second order logic.  They formalized our model and used a mechanized solver
to validate several examples.  In particular, they studied the following
program:
\begin{gather*}
  \tag{\ensuremath{P_6}} \label{eq:tar}
  \texttt{y=x; || z=1; || if(!z)\{x=y;\}\,else\,\{x=1;\}}
  \\
  \begin{tikzpicture}[node distance=1em,baselinecenter]
    \zevent{init}{\init}{}
    \zevent{Wz1}{\DW{z}{1}}{below=of init}
    \zevent{Rz1}{\DR{z}{0}}{below right=of init,xshift=8mm}
    \zevent{Rz0}{\DR{z}{1}}{right of=Rz1,node distance=7em}
    \zevent{0Wx1}{\DW{x}{1}}{below=of Rz0}
    \zevent{1Rx1}{\DR{y}{1}}{below right=of Rz1,xshift=-4mm}
    \zevent{1Rx0}{\DR{y}{0}}{below left=of Rz1,xshift=4mm}
    \zevent{1Wy0}{\DW{x}{0}}{below=of 1Rx0}
    \zevent{1Wy1}{\DW{x}{1}}{below=of 1Rx1}
    \zpo{init}{Rz0}
    \zpo{init}{Rz1}
    \zconfl{Rz0}{Rz1}
    \zconfl{1Rx0}{1Rx1}
    \zpo{Rz0}{0Wx1}
    \zpo{Rz1}{1Rx1}
    \zpo{Rz1}{1Rx0}
    \zpo{1Rx1}{1Wy1}
    \zpo{1Rx0}{1Wy0}
    \zpo{init}{Wz1}
    \zevent{Rx1}{\DR{x}{1}}{below left=of init,xshift=-8mm}
    \zevent{Rx0}{\DR{x}{0}}{left of=Rx1,node distance=5em}
    \zevent{Wy0}{\DW{y}{0}}{below=of Rx0}
    \zevent{Wy1}{\DW{y}{1}}{below=of Rx1}
    \zpo{init}{Rx0}
    \zpo{init}{Rx1}
    \zpo{Rx0}{Wy0}
    \zpo{Rx1}{Wy1}
    \zconfl{Rx0}{Rx1}
  \end{tikzpicture}
\end{gather*}
\citeauthor{pride} have established that there is no well-justified
configuration of \ref{eq:tar} that includes both $\DR{z}{0}$ and either
$\DR{x}{1}$ or $\DR{y}{1}$.  The rationale for forbidding this execution is
similar to that for TC5: \cite{PughWebsite} states that ``values are not
allowed to come out of thin air, even if there are other executions in which
the thin-air value would have been written to that variable by some not
out-of-thin air means.''

\section{Open problems}
\label{sec:open}  

As far as we are aware, it remains an open problem to define a sensible
semantics that allows the behavior of \ref{eq:tc9} from \S\ref{sec:outro}
while disallowing that of \ref{eq:tar} from \S\ref{sec:related}.  Our model
disallows \ref{eq:tc9}.  Speculative models that rely on purely existential reasoning allow
\ref{eq:tar}; this includes the models of \citet{DBLP:conf/popl/MansonPA05},
\citet{2010-gosrmm}, \citet{Kang:2017:PSR:3009837.3009850} and \citet{CV19}.

\emph{Coherence} is the property that writes on a single variable appear to
occur in some global order.  Our model does enforce coherence on
synchronization variables, where justification implies causal order, but not
on plain variables.  For example, in a coherent semantics for \texttt{(x=1;)
  || (x=2;) || (\zr1=x; \zr2=x; \zr3=x;)} it is not possible to have
$\zr1=\zr3\neq\zr2$, whereas our semantics allows this.  Our model is not
unique in this respect: This outcome is allowed by the JMM \citep{DBLP:conf/popl/MansonPA05} and
by the \emph{Local DRF} model of \citep{DBLP:conf/pldi/DolanSM18}; C++ gives
undefined behavior for this program.
%
To enforce coherence on plain variables, it appears to be necessary to
distinguish the causal order from the order used to determine visibility.


We have modeled synchronization using a restricted form of locks.  A coherent
semantics is required to model Java's volatile variables, and would also allow
a more satisfactory treatment of locks.  In the formalization of
\textsection\ref{sec:fencing}, release and acquire are both read/write
actions.  This guarantees that, for example, a single release does not enable
two parallel acquires.  In the standard treatment of locks, assuming
coherence, release is a write, and acquire is a read; thus, release justifies
acquire, but acquire does not justify release.  The order between acquire and
release is usually guaranteed by thread order, which we have assumed.  This
assumption guarantees that the order we have required from acquire to release
is redundant.  With a coherent semantics for locks, the causal order from
acquire to release can be dropped.



Separate from the concerns of \textsection\ref{sec:outro}, variations on the
definition of well-justification may be worth exploring.  For example, we
define of $\aeext$ in terms of $\acext$: when exploring extensions of the
current configuration, both player and opponent are restricted to using
acyclic justification.  It is natural to ask about a further definition,
which uses $\aeext$ in place of $\acext$.  If we let
$\mathord{\aeext_0}=\mathord{\acext}$ and
$\mathord{\aeext_1}=\mathord{\aeext}$, we can see this as hierarchy where
each $\aeext_i$ uses $\aeext_{i-1}$.  It is not the case that $\aeext_i$
contains $\aeext_{i-1}$, since the definition uses $\aeext_{i-1}$ in both
positive and negative position: positive for the player, and negative for the
opponent.  Example~\ref{ex:sevcik532} is interesting in this regard.  If the
opponent is allowed to pick using \AEjustification, then the player is unable
to well-justify the desired configuration.  Well-justification becomes
possible, however, if the opponent is restricted to acyclically chosen
configurations.

We have investigated a very simple program logic, which establishes a
restricted form of safety.  It would be interesting to investigate more
powerful logics, such as that of \citet{Turon:2014:GNW:2660193.2660243}.  
One of the primary purposes of a memory
model is to support program transformation.  To this end, it would be useful
to have a refinement relation over memory event structures that preserves
well-justification.  


Our approach to type safety is novel, in that we have not required a static
association between variables and types as in prior work
\citep{DBLP:journals/toplas/Lochbihler13,2010-relaxed-types}.  It would be
interesting to extend our approach to model dynamic allocation and
deallocation.

%% file: paper.bbl
\begin{thebibliography}{27}
\providecommand{\natexlab}[1]{#1}
\providecommand{\url}[1]{\texttt{#1}}
\expandafter\ifx\csname urlstyle\endcsname\relax
  \providecommand{\doi}[1]{doi: #1}\else
  \providecommand{\doi}{doi: \begingroup \urlstyle{rm}\Url}\fi

\bibitem[Alglave(2010)]{AlglaveThesis}
J.~Alglave.
\newblock \emph{A shared memory poetics}.
\newblock {PhD} thesis, Université Paris 7 and INRIA, 2010.

\bibitem[Alglave et~al.(2014)Alglave, Maranget, and Tautschnig]{AlglaveMT14}
J.~Alglave, L.~Maranget, and M.~Tautschnig.
\newblock Herding cats: Modelling, simulation, testing, and data mining for
  weak memory.
\newblock \emph{{ACM} Trans. Program. Lang. Syst.}, 36\penalty0 (2):\penalty0
  7:1--7:74, 2014.

\bibitem[Batty et~al.(2011)Batty, Owens, Sarkar, Sewell, and
  Weber]{DBLP:conf/popl/BattyOSSW11}
M.~Batty, S.~Owens, S.~Sarkar, P.~Sewell, and T.~Weber.
\newblock Mathematizing {C++} concurrency.
\newblock In \emph{Proceedings of the 38th ACM SIGPLAN-SIGACT Symposium on
  Principles of Programming Languages, POPL 2011, Austin, TX, USA, January
  26-28, 2011}, pages 55--66, 2011.

\bibitem[Batty et~al.(2015)Batty, Memarian, Nienhuis, Pichon{-}Pharabod, and
  Sewell]{DBLP:conf/esop/BattyMNPS15}
M.~Batty, K.~Memarian, K.~Nienhuis, J.~Pichon{-}Pharabod, and P.~Sewell.
\newblock The problem of programming language concurrency semantics.
\newblock In \emph{Programming Languages and Systems - 24th European Symposium
  on Programming, {ESOP} 2015, Held as Part of the European Joint Conferences
  on Theory and Practice of Software, {ETAPS} 2015, London, UK, April 11-18,
  2015. Proceedings}, pages 283--307, 2015.

\bibitem[Boehm and Adve(2008)]{DBLP:conf/pldi/BoehmA08}
H.-J. Boehm and S.~V. Adve.
\newblock Foundations of the {C++} concurrency memory model.
\newblock In \emph{Proceedings of the ACM SIGPLAN 2008 Conference on
  Programming Language Design and Implementation, Tucson, AZ, USA, June 7-13,
  2008}, pages 68--78, 2008.

\bibitem[Castellan(2016)]{Castellan}
S.~Castellan.
\newblock {Weak memory models using event structures}.
\newblock In \emph{{Vingt-septi{\`e}mes Journ{\'e}es Francophones des Langages
  Applicatifs (JFLA 2016)}}, Saint-Malo, France, Jan. 2016.
\newblock URL \url{https://hal.inria.fr/hal-01333582}.

\bibitem[Cenciarelli et~al.(2007)Cenciarelli, Knapp, and
  Sibilio]{DBLP:conf/esop/CenciarelliKS07}
P.~Cenciarelli, A.~Knapp, and E.~Sibilio.
\newblock The {Java} memory model: Operationally, denotationally,
  axiomatically.
\newblock In \emph{Programming Languages and Systems, 16th European Symposium
  on Programming, ESOP 2007, Held as Part of the Joint European Conferences on
  Theory and Practics of Software, ETAPS 2007, Braga, Portugal, March 24 -
  April 1, 2007, Proceedings}, pages 331--346, 2007.

\bibitem[Chakraborty and Vafeiadis(2017)]{DBLP:conf/cgo/ChakrabortyV17}
S.~Chakraborty and V.~Vafeiadis.
\newblock Formalizing the concurrency semantics of an {LLVM} fragment.
\newblock In \emph{Proceedings of the 2017 International Symposium on Code
  Generation and Optimization, {CGO} 2017, Austin, TX, USA, February 4-8,
  2017}, pages 100--110. {ACM}, 2017.

\bibitem[Chakraborty and Vafeiadis(2019)]{CV19}
S.~Chakraborty and V.~Vafeiadis.
\newblock Grounding thin-air reads with event structures.
\newblock \emph{{PACMPL}}, \penalty0 ({POPL}), 2019.
\newblock To Appear.

\bibitem[Cooksey et~al.(2018)Cooksey, Harris, Batty, Grigore, and
  Janota]{pride}
S.~Cooksey, S.~Harris, M.~Batty, R.~Grigore, and M.~Janota.
\newblock {PrideMM}: A solver for relaxed memory models.
\newblock \emph{CoRR}, abs/1901.00428, 2018.
\newblock URL \url{http://arxiv.org/abs/1901.00428}.

\bibitem[Dolan et~al.(2018)Dolan, Sivaramakrishnan, and
  Madhavapeddy]{DBLP:conf/pldi/DolanSM18}
S.~Dolan, K.~C. Sivaramakrishnan, and A.~Madhavapeddy.
\newblock Bounding data races in space and time.
\newblock In \emph{Proceedings of the 39th {ACM} {SIGPLAN} Conference on
  Programming Language Design and Implementation, {PLDI} 2018, Philadelphia,
  PA, USA, June 18-22, 2018}, pages 242--255. {ACM}, 2018.

\bibitem[Goto et~al.(2012)Goto, Jagadeesan, Pitcher, and
  Riely]{2010-relaxed-types}
M.~Goto, R.~Jagadeesan, C.~Pitcher, and J.~Riely.
\newblock Types for relaxed memory models.
\newblock In \emph{Proceedings of {TLDI} 2012: The Seventh {ACM} {SIGPLAN}
  Workshop on Types in Languages Design and Implementation, Philadelphia, PA,
  USA, Saturday, January 28, 2012}, pages 25--38. {ACM}, 2012.

\bibitem[Jagadeesan et~al.(2010)Jagadeesan, Pitcher, and Riely]{2010-gosrmm}
R.~Jagadeesan, C.~Pitcher, and J.~Riely.
\newblock Generative operational semantics for relaxed memory models.
\newblock In \emph{Programming Languages and Systems, 19th European Symposium
  on Programming, {ESOP} 2010, Held as Part of the Joint European Conferences
  on Theory and Practice of Software, {ETAPS} 2010, Paphos, Cyprus, March
  20-28, 2010. Proceedings}, volume 6012 of \emph{Lecture Notes in Computer
  Science}, pages 307--326. Springer, 2010.

\bibitem[Jeffrey and Riely(2014)]{talk14}
A.~Jeffrey and J.~Riely.
\newblock Event structures and refinement for relaxed memory.
\newblock Presentation at the Memory Model Meeting, Cambridge, UK, Sept. 2014.
\newblock URL \url{https://perma.cc/BG4K-HSVM}.

\bibitem[Kang et~al.(2017)Kang, Hur, Lahav, Vafeiadis, and
  Dreyer]{Kang:2017:PSR:3009837.3009850}
J.~Kang, C.-K. Hur, O.~Lahav, V.~Vafeiadis, and D.~Dreyer.
\newblock A promising semantics for relaxed-memory concurrency.
\newblock In \emph{Proceedings of the 44th ACM SIGPLAN Symposium on Principles
  of Programming Languages}, POPL 2017, pages 175--189, New York, NY, USA,
  2017. ACM.

\bibitem[Lamport(1979)]{Lamport}
L.~Lamport.
\newblock How to make a multiprocessor computer that correctly executes
  multiprocess programs.
\newblock \emph{IEEE Trans. Comput.}, 28\penalty0 (9):\penalty0 690--691, 1979.

\bibitem[Lochbihler(2013)]{DBLP:journals/toplas/Lochbihler13}
A.~Lochbihler.
\newblock Making the java memory model safe.
\newblock \emph{{ACM} Trans. Program. Lang. Syst.}, 35\penalty0 (4):\penalty0
  12, 2013.

\bibitem[Manson et~al.(2005)Manson, Pugh, and Adve]{DBLP:conf/popl/MansonPA05}
J.~Manson, W.~Pugh, and S.~V. Adve.
\newblock The {Java} memory model.
\newblock In \emph{Proceedings of the 32nd ACM SIGPLAN-SIGACT Symposium on
  Principles of Programming Languages, POPL 2005, Long Beach, California, USA,
  January 12-14, 2005}, pages 378--391, 2005.

\bibitem[Nielsen et~al.(1979)Nielsen, Plotkin, and
  Winskel]{Nielsen:1979:PNE:647172.716120}
M.~Nielsen, G.~D. Plotkin, and G.~Winskel.
\newblock Petri nets, event structures and domains.
\newblock In \emph{Semantics of Concurrent Computation, Proceedings of the
  International Symposium, Evian, France, July 2-4, 1979}, volume~70 of
  \emph{Lecture Notes in Computer Science}, pages 266--284. Springer, 1979.

\bibitem[Pichon{-}Pharabod and Sewell(2016)]{Sewell16}
J.~Pichon{-}Pharabod and P.~Sewell.
\newblock A concurrency semantics for relaxed atomics that permits optimisation
  and avoids thin-air executions.
\newblock In \emph{Proceedings of the 43rd Annual {ACM} {SIGPLAN-SIGACT}
  Symposium on Principles of Programming Languages, {POPL} 2016, St.
  Petersburg, FL, USA, January 20 - 22, 2016}, pages 622--633, 2016.

\bibitem[Pugh(2004)]{PughWebsite}
W.~Pugh.
\newblock Causality test cases, 2004.
\newblock URL \url{https://perma.cc/PJT9-XS8Z}.

\bibitem[Saraswat et~al.(2007)Saraswat, Jagadeesan, Michael, and von
  Praun]{Saraswat:2007:TMM:1229428.1229469}
V.~A. Saraswat, R.~Jagadeesan, M.~Michael, and C.~von Praun.
\newblock A theory of memory models.
\newblock In \emph{Proceedings of the 12th ACM SIGPLAN Symposium on Principles
  and Practice of Parallel Programming}, PPoPP '07, pages 161--172, 2007.

\bibitem[Sezgin(2014)]{rng}
A.~Sezgin.
\newblock Stress testing.
\newblock Private correspondence, Oct. 2014.

\bibitem[Turon et~al.(2014)Turon, Vafeiadis, and
  Dreyer]{Turon:2014:GNW:2660193.2660243}
A.~Turon, V.~Vafeiadis, and D.~Dreyer.
\newblock {GPS}: Navigating weak memory with ghosts, protocols, and separation.
\newblock In \emph{Proceedings of the 2014 ACM International Conference on
  Object Oriented Programming Systems Languages \& Applications}, OOPSLA '14,
  pages 691--707, 2014.

\bibitem[\v{S}ev\v{c}\'{\i}k(2008)]{SevcikThesis}
J.~\v{S}ev\v{c}\'{\i}k.
\newblock \emph{Program Transformations in Weak Memory Models}.
\newblock {PhD} thesis, Laboratory for Foundations of Computer Science,
  University of Edinburgh, 2008.

\bibitem[Winskel(1986)]{DBLP:conf/ac/Winskel86}
G.~Winskel.
\newblock Event structures.
\newblock In \emph{Petri Nets: Central Models and Their Properties, Advances in
  Petri Nets 1986, Part II, Proceedings of an Advanced Course, Bad Honnef,
  Germany, 8-19 September 1986}, volume 255 of \emph{Lecture Notes in Computer
  Science}, pages 325--392. Springer, 1986.

\bibitem[Winskel(1988)]{Winskel:1988:IES:648140.749805}
G.~Winskel.
\newblock An introduction to event structures.
\newblock In \emph{Linear Time, Branching Time and Partial Order in Logics and
  Models for Concurrency, School/Workshop, Noordwijkerhout, The Netherlands,
  May 30 - June 3, 1988, Proceedings}, volume 354 of \emph{Lecture Notes in
  Computer Science}, pages 364--397. Springer, 1988.

\end{thebibliography}
